\documentstyle[epsfig, psfrag,12pt,eqn]{article}
\slshape
\newcommand{\be}{\begin{equation}}
\newcommand{\ee}{\end{equation}}
\newcommand{\bef}{\begin{displaymath}}
\newcommand{\eef}{\end{displaymath}}
\newcommand{\bes}{\begin{eqnarray}}
\newcommand{\ees}{\end{eqnarray}}
\newcommand{\besf}{\begin{eqnarray*}}
\newcommand{\eesf}{\end{eqnarray*}}

\newcommand{\margen}{\hspace{8mm}}

\begin{document}

\title{THE PRIMORDIAL \\ GRAVITATIONAL WAVE BACKGROUND \\
IN  STRING COSMOLOGY}

\author{M.P.Infante$^{(1),(2)}$\footnote{E-mail: infante@mesunb.obspm.fr}, N.S\'anchez$^{(1)}$\footnote{E-mail: Norma.Sanchez@obspm.fr}   \\
{\footnotesize {\it (1) Observatoire de Paris-DEMIRM. 61, Avenue de 
l'Observatoire, 75014 Paris, FRANCE }} \\
{\footnotesize {\it (2) Dpt. F\'{\i}sica Te\'orica, Univ. de Zaragoza. Pza.San Francisco, 50009 Zaragoza, SPAIN}}}
\date{\empty}
\maketitle
\vspace{-1cm}

\begin{abstract}

We find the spectrum $P(\omega)d\omega$ of the gravitational wave
background produced in the early universe in string theory. 
We work in the framework of the recently discussed
String Driven Cosmology, 
whose scale factors are computed with the low-energy effective string equations
as well as selfconsistent solutions of General Relativity with a gas 
of strings as source.
The scale factor evolution is described by an early string driven
inflationary stage with an instantaneous transition to a
radiation dominated stage and successive matter dominated stage.
This is a string cosmology expanding evolution always running on positive
proper cosmic time.

The spectrum of the generated gravitons is computed
in the framework of Quantum Field Theory
and in the appropiated effective String Cosmology context. 
We study and show explicitly the effect of the dilaton field, 
characteristic to
this kind of cosmologies.
We compute the spectrum for the same evolution description with
three differents approachs. 
Some features of gravitational wave spectra, as peaks and asymptotic
behaviours, are found direct consequences of the dilaton involved in 
the string low energy effective action and not only of the scale factor 
evolution.

We make use of a careful treatment of the scale factor evolution and
involved transitions. This allows us to compute a full 
prediction on the power spectrum of gravitational waves without any
free-parameters.  A comparative 
analysis of different treatments, solutions and compatibility
with observational bounds or detection perspectives is made. 

{\bf PACS number(s):} 98.80.Cq; 11.25.-w; 04.30.Db; 04.80.Cc

{\bf Report numbers:} hep-th/9907185; DFTUZ/99/13.
\end{abstract}

\section{Introduction and Results}
\setcounter{equation}0

{\margen} In the latest years, an impressive scientific effort has been made 
on the subject of production of relic gravitational waves.
The spectrum of relic gravitational waves constitutes an important source of 
information about the conditions and dynamics of the early Universe. 
From the pioneering works relating the very early Universe with the
spectrum of metric fluctuations (\cite{gkz},\cite{st},\cite{fpa}) 
a lot of work has followed computing the gravitational waves background
in different evolution models (see for instance
\cite{A},\cite{gks},\cite{gkprd},\cite{abha},\cite{Sah},\cite{wh}).
The relic gravitational wave background is generated as a consequence 
of the amplification by the Universe expansion of the graviton fields, which 
decouples from matter at very early stages. Therefore, their spectrum is 
defined by conditions in the very early epoch and remains practically 
unaffected until our days, since the medium is transparent to their 
propagation. 

In addition, important improvements on the projects for direct detection of 
gravitational waves, as well as on the measurements of astrophysical phenomena
affected by them have been performed or are expected for the next decade. 
This will make possible to extract constraints and to put upper limits on the 
magnitude of the generation and amplification of gravitational waves and 
thereafter, on the particular model of evolution considered for the universe. 
Currently, the search of direct measurements by means of resonant bar 
detectors (e.g., NAUTILUS, AURIGA) has been followed by the development of 
laser interferometers (LIGO, LISA). The sensitivities to be reached for the 
next generation of laser interferometers detectors could fix these constraints
on ranges still more interesting from the point of view of early universe 
theories (\cite{ab}).
Indirect constraints can be obtained from the closure relation, 
nucleosynthesis bound, pulsar timing measurements and isotropy of Cosmic 
Microwave Background Radiation (CMB) (\cite{lnr}). Finally, there
would be imprints on the CMB temperature anisotropy spectrum through 
Sachs-Wolfe effect (\cite{sw}) and polarization fluctuations originated 
by scattering 
Thompson on a tensorially perturbed metric (\cite{hw}).

This study is particularly interesting from the point of view of String Theory
 in Cosmology since gravitational waves are one of the observable tracks to be 
extracted from the fundamental theory. The decoupling of gravitons from 
another kind of matter has taken place very close to the Planck Energy Scale. 
Thus, the shape of the spectrum could carry information about the physics on 
such energy and lenght scales. A number of studies computing 
gravitational wave production in the context of
string cosmology scenarii
has been made in the last years. 
(\cite{gv93},\cite{gdil},\cite{bggv},\cite{buon}). Oughted to the intrinsic 
uncertainty in the cosmological model, the spectra computed in
these works depend on free parameters coming from the earliest
stages of Universe, far from observational knowledge. No firm
free-parameter prediction is extracted, although the existence of a peak 
and end-point in the string cosmology gravitational wave background
has been claimed (\cite{bgv}).  

In this paper,  we study the production of a primordial stochastic
gravitational wave background generated in the context of selfconsistent 
string cosmology (\cite{dvs94},\cite{lib1},\cite{lib3},\cite{lib2}). 
From both treatments therein 
explained, low energy effective string action and selfconsistent solutions
of General Relativity plus string matter, we have obtained a minimal
model for the evolution of the scale factor \cite{p2}. 
The String Driven Cosmological
Background is characterized by an inflationary inverse power stage 
$a_I(t) \sim {(t_I-t)}^{-\frac{2}{d+1}}$
plus a radiation dominated stage 
$a_{II}(t) \sim  {(t-t_{II})}^{\frac{2}{d+1}}$
and a matter dominated stage 
$a_{III}(t) \sim t^{\frac{2}{d}}$. 
The cosmological model is linked to the Universe observational
information by means of descriptive temporal variables as performed 
in \cite{p1} 
and \cite{p2}. Our results are thus expressed in terms of standard times 
for inflation-radiation dominated
transition ${\mathcal{T}}_r$, and radiation dominated-matter dominated
transition ${\mathcal{T}}_m$.  The conformal time variable $\eta$ has
been defined with totally satisfactory continuity conditions in the
transitions.

We compute exact expressions for the power spectrum $P(\omega) d\omega$
and contribution to the energy density $\Omega_{GW}$ of primordial 
gravitational waves
generated in the transition among the inflationary and radiation dominated
stages. These expressions are fully predictive and free-parameter, since the
cosmological model has been constrainted with minimal and well stablished
observational Universe information (the transition times ${\mathcal{T}}_r$ and 
${\mathcal{T}}_m$). Having fixed a minimal but satisfactory model for the
scale factor, we study the production of gravitational waves by
making three approaches in the perturbation treatment.

The role of the dilaton, a characteristic field in low energy effective 
string theory and present in our
model, is proved to be crucial since gives a strong signature on the shape
and magnitude of these spectra. The computations of graviton perturbations
with
no dilaton role are showed to be equivalent to standard Quantum Field
Theory computations, like those reported in Allen {\cite{A}} and Grishchuk
{\cite{gks}}, but notorious differences are found in the spectra
oughted to the particular character of the inflationary
string driven scale factor, an inverse power type. We obtain an exact
expression for the
power spectrum and energy density contribution (eqs.(\ref{P:All})
and (\ref{om:All})) in
terms of Hankel functions. The power spectrum asymptotic behaviours at
low and high frequencies are both vanishing with dependences
$\omega^{\frac{1}{3}}$ and ${\omega}^{-1}$ respectively. 
This gives a gravitational wave contribution to the energy density 
asymptotically constant at high
frequencies on $\Omega_{GW} \sim 10^{-26}$.
The slope change
produces a peak in the power spectra around a characteristic frequency totally
determined by the model of value
$\omega_x \; \sim  \: 1.48 \; {\mathrm{Mhz}}$.

Computations with partial dilaton role, considering this in the
perturbation equation but not on the perturbation itself,
show some differences with the last ones, althought the general
characteristics are very similar. The formal expressions for the
power spectrum and energy density contribution are the same of
the No Dilaton case (eqs.(\ref{P:All}) and (\ref{om:All})), but 
they involve Hankel functions with different orders. Both 
asymptotic regimes for small and high frequencies vanish again, but now with 
dependences ${\omega}^{\frac{5}{3}}$ and ${\omega}^{-1}$ respectively.
The peak appears around the same characteristic frequency,
but its value is one order of magnitude lower than in the
No Dilaton case, as well as the asymptotic constant contribution
to the energy density $\Omega_{GW}$.

In the opposite case, when the full
dilaton role is accounted, general characteristics as well as orders
of magnitude of the graviton spectrum are drastically modified. 
The formal expressions are different (eqs.(\ref{P:FD}),(\ref{om:FD})).
We extract the asymptotic behaviours too: for low frequencies, the
dependence is vanishing again with $\omega^{\frac{5}{3}}$. For
high frequencies, an asymptotic 
divergent behaviour proportional to $\omega$ is found. Similarly, the 
contribution to the
critical energy density in gravitational waves is
divergent at high frequencies, as $\omega^{2}$. 
The change of slope is
less visible and no clear peaks are found. The transition
from the low frequency to the high frequency regime is
slower than in the previous cases (with no dilaton or partial
dilaton roles) and full analytical
expressions are needed on a wider range 
$10^6 \sim 10^9 Hz$. 
Comparaison of these summarized results can be
found in Tables (\ref{spectra}), (\ref{fracs}) and (\ref{limi})
and in Figures (\ref{f.esp}) and (\ref{f.ome}).

The behaviour obtained in the full dilaton treatment is a
common characteristic to the String Cosmology low
energy effective treatments. We conclude that the
Brans-Dicke frame (dilaton field) involved and the
type of scale factor predicted (inverse power type)
are responsible of these features.
More conclusions (the dilaton role, string and no-string
cosmologies) are given in Section VII.

In Section II, we recall the derivation of the background solutions
obtained in String Cosmology, the richness of String Theory 
applied to cosmology is visible along this section. In Section
III we summarize the String Driven
Cosmological Background in order to obtain an
approximative but suitable description of the cosmology of our
observational Universe. We
translate this description to a conformal time-type variable more
convenient for further work. Section IV shows the computation of
gravitational wave production in the framework of Quantum Field
Theory and shows how the
dilaton role can be considered. In Section V we particularize to 
our cosmological description and obtain exact expressions in 
suitable measurement units both for power spectrum and
energy density in each one of the three treatments.  In Section VI 
we analyze the 
asymptotic behaviours. Finally, we elaborate and discuss
our results and compare to other String Cosmology computations as well
as to pure General Relativity
gravitational wave computations, and present our 
conclusions in Section VII.

\section{String Driven Cosmological Background}
\setcounter{equation}0

{\margen} The String Driven Cosmological Background is a minimal
model of the Universe evolution totally extracted from effective
String Theory. Details on the model can be found in Ref.\cite{p2}, but
here we summarize it. Two ways allowing  extraction of  cosmological
descriptions for the background from string theory have
been used. One is 
the low energy effective string equations plus string action matter.
Solutions of these equations are 
an inflationary inverse power law evolution of the scale factor, as well
as a radiation dominated behaviour. On the other hand, selfconsistent Einstein
equations with a classical gas of strings as sources will give us
again a radiation dominated behaviour and a matter dominated description.

In this whole and next sections, unless opposite indication, the
metric is defined in lenght units. Thus, the (0,0) component
is always time coordinate ${\mathcal{T}}$ multiplied by constant $c$, 
that means  $t = c {\mathcal{T}}$.
Derivatives are taken with respect to this coordinate $t$. Otherwise,
constants $c$, $\hbar$ and $G$ are explicitly showed.
We consider a spatially flat, homogeneous and isotropic background and we 
write the metric 
in synchronous frame ($g_{00} =1$, $g_{0i}=0=g_{0a}$) as:
\be\label{metrica}
g_{ \mu \nu}={\mathrm{diag}}(1, -a^2(t) \; {{\delta}_i}_j)
\ee

The earliest stages in our model are provided by the low energy effective 
context. The scale factor is extracted by extremizing 
with respect to dilaton field, graviton field and matter sources  the low 
energy string 
effective action (to the lowest order in expansion of powers of $\alpha'$), 
which in the Brans-Dicke or string frame can be written as 
\cite{tsy},\cite{dvs94},\cite{gv93},\cite{lib3}:
\be \label{action}
S = - \frac{c^3}{16 \pi G_D} \int d^{d+1} x \sqrt{\mid g \mid} e^{- \phi}
\left(R + \partial_{\mu} \phi \partial^{\mu} \phi - \frac{H^2}{12} + V \right)
+ S_M
\ee
where $S_M$ is the corresponding action for the strings as matter sources, 
$H=dB$ is the 
antisymmetric tensor field strenght and $V$ is related with dilaton potential
and vanishing for some 
critical dimension.  The dilaton field $\phi$ depends explicitly only 
upon time coordinate. $D$ is the
total spacetime dimension, $D=d+1$ where $d$ is the number of spatial
dimensions.
The string matter is included as a classical source which stress
energy tensor in the perfect fluid approximation takes the form:
\be \label{source}
{T_{\mu}}^{\nu} = {\mathrm{diag}}(\rho(t), -P(t) {{\delta}_i}^j)
\ee
where $\rho$ is the energy density and $P$ the pressure for the matter sources.
Here, we will consider antisymmetric tensor $H_{\mu \alpha
\beta}$ as well as the potential $B$ vanishing. We do not consider here
the effects oughted to a non-vanishing dilaton potential. Thus, 
the low energy effective equations are obtained:
\pagebreak[4]
\bes \label{leeeq}
{\dot{\bar{\phi}}}^{\:2} - 2 {\ddot{\bar{\phi}}} + d H^2 & = & 0 \\
{\dot{\bar{\phi}}}^{\:2} - d H^2 & = & \frac{16 \pi G_D}{c^4} \; {\bar{\rho}} 
\; e^{\bar{\phi}} \nonumber \\
2 ({\dot{H} - H {\dot{\bar{\phi}}}}) & = & \frac{16 \pi G_D}{c^4} \; {\bar{p}} 
\; e^{\bar{\phi}} \nonumber
\ees
where $H = {{\dot{a}}\over{a}}$ and we use the shifted string
duality invariant expressions for
the dilaton $\bar{\phi}=\phi-\ln{\sqrt{\mid g \mid}}$, matter energy
density $\bar{\rho} =\rho a^d$ and pressure $\bar{p}=P a^d$.

The inflationary String Driven stage appears as a new selfconsistent 
solution of the low energy effective
equations (\ref{leeeq}) sustained by a gas of stretched or unstable string
sources (\cite{dvs94},\cite{lib1}). The pressure and energy
density of this string behaviour in curved background satisfy 
the equation of state $P = \gamma \rho$ as matter sources, 
where $d \; \gamma = -1$ for the unstable behaviour in the
metric here considered.
Thus, the equation of state to be considered for the string sources is 
given by:
\be \label{eqssd}
P = - \frac{1}{d} \ \rho
\ee
With this, the selfconsistent solution of effective 
equations (\ref{leeeq}) is:
\bes \label{sdrin}
a(t) & = & A_I ({t_I - t})^{-Q} \ \ \ \ \ \ 0 < t < t_r < t_I \ \ \ \ \ , 
\ \   
Q = \frac{2}{d+1}  \\
\phi(t) & = & \phi_I + 2d \ln a(t)  \nonumber \\
\rho(t) & = & \rho_I {(a(t))}^{(1-d)}  \nonumber \\
P(t) & = & -{1\over{d}} \; \rho(t) \ \ = \ \ -\frac{\rho_I}{d} 
{(a(t))}^{(1-d)} \nonumber
\ees
Notice here $t$ is the proper cosmic time coordinate, running on positive 
values; 
$d$ is the number of expanding spatial dimensions; $\rho_I$, $\phi_I$ are
integration constants and $A_I$, $t_I$ parameters to be fixed by
the further evolution of scale factor, the parameter $t_I$ being greater 
than the exit of the inflationary stage $t_r$.

It is convenient for further work to express the solutions in conformal 
time $\eta$ such $d \eta = {{d t}\over{a(t)}}$:
\bes \label{Sdrin}
a_{inf}(\eta) & \sim &  (\eta_I - \eta)^q \ \ \ \ \ , \  \ \ \ \ \  
q = {-\frac{2}{d+3}} \\
\phi_{inf}(\eta) & \sim & - {{4 d} \over{d+3}}\ln (\eta_I - \eta) \nonumber
\ees
where $ \eta_I $ is a parameter to be determined in 
the matching with the next cosmological stage.

As discussed in ref.\cite{p2}, this is a new string cosmology solution,
obtained without exploting manifestly duality relations among the
solutions, and describing
an accelerated expansion running
on the positive branch of cosmic temporal coordinate. In this way, this
solution is different to the ``Pre-Big Bang'' solutions
found in literature \cite{gv93}.

On the other hand, the dual to unstable string behaviour in
curved backgrounds follows a 
typical radiation type equation of state 
$P = \frac{1}{d} \  \rho $ \cite{dvs94}. The effective equations (\ref{leeeq})
plus this equation of state give the following behaviour
describing a radiation dominated stage \cite{dvs94}, \cite{gv93}:
\bes \label{sdrad}
a(t) & = & A_{II}\: t^R \ \ \ \ \ t_r < t  \ \ \  , \ \ \  \ \ 
R = \frac{2}{d+1}  \\
\phi(t) & = & \phi_{II} \nonumber \\
\rho(t) & = & \rho_{II} {(a(t))}^{-(1+d)} \nonumber \\
P(t) & = & \frac{1}{d} \; \rho(t) = \frac{\rho_{II}}{d}{(a(t))}^{-(1+d)} 
\nonumber
\ees
here $\phi_{II}$, $\rho_{II}$ are integration constants, and $A_{II}$ a 
parameter to be fixed by the evolution of scale factor. Notice 
the dilaton remains ``frozen'' at constant value. Again, 
we define the corresponding conformal time to this stage. The solution 
takes the form:
\bes \label{Sdrad}
a_{rad} (\eta) & \sim & (\eta - \eta_{II})^r \ \ \ , \  \ \  
r = \frac{2}{d-1} \\
\phi_{rad}(\eta) & = & \phi_{II} \nonumber
\ees
where $\eta_{II}$, $\phi_{II}$ are constants to be determined with the
scale factor evolution.

Cosmological backgrounds can be found too,
as selfconsistent solutions of the General Relativity Einstein equations 
selfsustained by the string sources evolving in them, as
developped in \cite{dvs94},\cite{lib1},\cite{lib3}. In the curved 
backgrounds here considered,
the Einstein equations take the form:
\bes \label{eins}
{1\over{2}}d(d-1)H^2 & =  & \rho \\
(d-1)\dot{H} + P + \rho & = & 0 \nonumber
\ees
and the matter source is described by a gas of non interacting 
classical strings, whose equation of state includes the different 
possible behaviours of strings in curved spacetimes: unstable, 
dual to unstable and stable, each one with string densities  
$\mathcal{U}$, $\mathcal{D}$ and $\mathcal{S}$ respectively.
Taking into account the properties of each behaviour (see \cite{dvs94}), the 
energy density and the pressure of the string gas are described by: 
\bes \label{gasro}
\rho & =  & \frac{1}{{(a(t))}^d} \left({\mathcal{U}} a(t) + 
{{\mathcal{D}}\over
a(t)} + \mathcal{S}\right) \\
P & = & \frac{1}{d} \; \frac{1}{{(a(t))}^d} 
\left( {{\mathcal{D}}\over a(t)} -
{\mathcal{U}} a(t) \right) \label{gasp}
\ees
Equations (\ref{gasro}) and (\ref{gasp}) are qualitatively corrects for every 
$t$ and become exacts in the asymptotic regimes, leading to obtain the 
radiation 
dominated behaviour of the scale factor, as well as 
the matter dominated behaviour. The first one is obtained in the limit
$a(t)\rightarrow 0$ and $t \rightarrow 0$, where the dual to unstable 
behaviour dominates in the equations (\ref{gasro}) and (\ref{gasp}) and 
gives us:
\bes \label{rdrp}
\rho(t) & \sim & {\mathcal{D}} \; {(a(t))}^{-(d+1)} \\
P(t) & \sim & \frac{1}{d} \; {\mathcal{D}} \; {(a(t))}^{-(d+1)} \nonumber 
\ees
Solving the Einstein equations (\ref{eins}) with sources following 
eqs.(\ref{rdrp}), yields the scale factor solution:
\be \label{dv}
a(t) \sim  {\left({{2 \mathcal{D}}\over{d(d-1)}}\right)}^{1\over{d+1}} 
{\left({{d+1}\over{2}}\right)}^{2\over{d+1}} (t - t_{II})^R \ \ \ \ \ ,  \ \ 
 R = \frac{2}{d+1}
\ee
$a(t)$ describes the evolution of a FRW radiation
dominated stage, here the parameter $t_{II}$ will be fixed by
the further evolution of the scale factor. This is coherent with
the fact that dual to unstable strings behave in a similar way to 
massless particles, i.e. radiation \cite{dvs94}.
In conformal time and rearranging constants,  this solution reads:
\bef
a_{rad}(\eta)  =  \alpha_{II} (\eta - \eta_{II})^r 
\ \ \ \ \ ,  \ \ \    r =\frac{2}{d-1}
\eef
Notice that the scale factor now described is equivalent to the
before
obtained in low energy effective treatment, eq.(\ref{Sdrad}).

In the opposite limit $a(t)\rightarrow 
\infty $, $t \rightarrow \infty$,  the unstable density ${\mathcal{U}} 
\rightarrow  0 $ and the stable behaviour ${\mathcal{S}}$
becomes dominant. The 
equation of state reduces to:
\bes \label{mdrp}
\rho & \sim & {\mathcal{S}} \; {(a(t))}^{-d} \\
P & = & 0 \nonumber
\ees 
and from solving eqs.(\ref{eins}) with 
eqs.(\ref{mdrp}), the solution of a matter dominated stage emerges
(Notice that stable strings behave as
cold matter):
\be \label{rgmat}
a(t) \sim  {\left({d\over{(d-1)}}{{\mathcal{S}}\over{2}}\right)}^{1\over{d}} 
(t-t_{III})^M \ \ \  ,  \ \ \ \  M = \frac{2}{d}
\ee
here $t_{III}$ is fixed by the whole scale factor evolution. This scale factor
in conformal time takes the form:
\be \label{Rgmat}
a_{mat}(\eta)  =  \alpha_{III} (\eta - \eta_{III})^m
 \  \  \  , \  \  \    m = \frac{2}{d-2}
\ee
here the constants $\alpha_{III}$ and $\eta_{III}$ will be determined 
from the transition 
between radiation and matter dominated stages.

The evolution model with the String Driven inflationary stage 
eq.(\ref{sdrin}), followed by a radiation
dominated stage (eqs.(\ref{sdrad}) or (\ref{dv}) are equivalents for our
purposes) and a matter dominated stage eq.(\ref{rgmat}) is the String 
Driven Cosmological Background that we will study.

Notice that we consider the dilaton field remaining  
practically constant and vanishing from the exit of inflation, as suggested
in the String Driven Radiation Dominated Solution, until the current time.
Notice also the Brans-Dicke frame for the metric-dilaton coupling 
where the string action has been written eq.(\ref{action}) and
both an inflationary and a radiation dominated stages have been
extracted. The last one
can be obtained too selfconsistently from the Einstein equations 
plus string matter, the same treatment where the matter dominated current
stage is obtained too.
In the table (\ref{solut}) the scale factor, dilaton and equation of state 
obtained from these treatments can be observed in a comparative way.

\begin{table}[hp]
\centering 
\begin{tabular}{||c||c|c||c|c||}
\hline
\hline
\empty & \multicolumn{2}{c||}{{\emph{String Driven Solutions}}}
& \multicolumn{2}{c||}{{\emph{String Sources in G.R.}}} \\ 
\cline{2-5}
\empty & \multicolumn{2}{c||}
{{LEE + (${\mathcal{U}}$,${\mathcal{D}}$) strings}}&
\multicolumn{2}{c||} {EE + 
gas (${\mathcal{U}}$,${\mathcal{D}}$,${\mathcal{S}}$) strings} \\ 
\cline{2-5}
\empty & inflation & radiation & radiation & matter \\ \hline
{$ a(t) \sim $} & {$ {(t_I - t)}^{-{2\over{d+1}}} $} & {$t^{2\over{d+1}}$}
& {$ t^{2\over{d+1}}$} & {$ t^{2\over{d}} $} \\
{$ a(\eta) \sim $} & {$(\eta_I - \eta)^{-{2\over{d+3}}}$} & {$(\eta - 
\eta_{II})^{2\over{d-1}}$} & {$ (\eta - \eta_{II})^{2\over{d-1}}$} & {$ (\eta -
\eta_{III})^{2\over{d-2}} $} \\
{$ \phi(t) \sim $} & {$\left({-{4d \over{d+1}}}\right) \ln (t_I -t)$} & 
cte  & - -  & - -  \\
{$ \phi(\eta) \sim $} & {$\left(-{4d\over{d+3}}\right)\ln (\eta_I - \eta)$} & 
cte  & - - & - - \\
{$ \rho(t) \sim $} & {${(a(t))^{(1-d)}}$} & {${(a(t))}^{-(1+d)}$} &
{$ {\mathcal{D}} {(a(t))}^{-(1+d)} $} & {${\mathcal{S}} {(a(t))}^{-d} $} \\
{\it string} & {\it unstable} & {\it dual to u.} & {\it dual to u.} & 
{\it stable} \\
{\it sources} & {$ P=-{1\over{d}}\rho $}  & {$ P={1\over{d}}\rho $}  & 
{$ P={1\over{d}} \rho $} & $P = 0$ \\
\hline \hline
\end{tabular}
\vspace*{5pt}
\caption{\label{solut} Selfconsistent solutions giving the String Driven 
Cosmological Background}
\vspace{5pt}
{\parbox{135mm}{\footnotesize Table comparative of the solutions for the scale
 factor $a(t)$, $a(\eta)$. $t$ is cosmic time, $\eta$ is conformal time and 
$\phi$ is the dilaton field. $LEE$ means low energy string 
effective equations in the Brans-Dicke frame, for a $(d+1)$ homogeneus, 
isotropic
and spatially flat background; antisymmetric tensors and dilaton 
potentials are neglected. $EE$ means the General Relativity Einstein
Equations plus a gas of strings as classical sources matter. 
The dominant string behaviour as matter sources and their corresponding 
equation of state are written in the last row. }} 
\end{table}

\section{Scale Factor Description}
\setcounter{equation}0

{\margen} Since the behaviours above extracted are asymptotic results, it is
not possible to give here the detail of transitions among the different stages.
It would be expected some changes in the regimes or equations
governing each stage and leading to the next one, but this is an open 
question in the framework of string theory both for inflation-radiation 
dominated as well as  radiation dominated-matter dominated transition.
Taking the simplest option, we consider the ``real'' scale factor evolution
minimally described as:
\bes \label{real}
a_I(t) & = & A_I {(t_I - t)}^{-Q} \ \ \ \   t \in (t_i, t_r)  \\
a_{II}(t) & = & A_{II} \: t^R \ \ \ \ \ \ \ \ \  t \in (t_r, t_m) \nonumber \\
a_{III}(t) & = & A_{III} \: t^M \ \ \ \ \ \ \ \ \ t \in (t_m, t_0) \nonumber
\ees
with transitions at least not excessively long at $t_r$ beginning of radiation 
dominated stage and $t_m$ beginning  of matter dominated stage. We define 
also a beginning of inflation at $t_i$ and a current  time $t_0$. 

From the point of view of String Theory applied to Cosmology, it would 
be reasonable to have not instantaneous and continuous transitions at
$t_r$ and $t_m$ \cite{p2}. But this condition greatly difficults the 
computation that we want make because it introduces free parameters, 
loss of physical meaning and effects of an unknown transition dynamics
\cite{A} oughted to discontinuity on the scale factor.
In order to construct a suitable minimal model for gravitational wave
computations, we will merge our lack of knowledge on real transitions 
by means of descriptive temporal variables in function of which the modelized 
transitions at $\bar{t_1}$ and $\bar{t_2}$ are instantaneous and continuous. 

We link this descriptive scale factor with the minimal information
about the observational Universe. We consider the standard values for 
cosmological times: the radiation-matter transition held about 
${\mathcal{T}}_m \sim 10^{12} s$, the beginning of radiation stage at
${\mathcal{T}}_r \sim 10^{-32} s $ and the current age of the Universe
${\mathcal{T}}_0 \sim {H_0}^{-1} \sim 10^{17} s $ (The exact numerical
value of ${\mathcal{T}}_0$ turns out not crucial here). We impose also
to our description to satisfy the same
scale factor expansion (or scale factor ratii) reached in each one of
the three stages considered in the real model (\ref{real}).
Explicit computations can be found in \cite{p1} and leds finally to
this scale factor written in cosmic time-type variables in the more
convenient way for our further work:
\bes \label{Descr}
\bar{\bar{a_I}}(\bar{\bar{t}}) & = & \bar{\bar{A_{I}}}
{(\bar{\bar{t_I}}-\bar{\bar{t}})}^{-Q} \ \ \ \ \ \ \ 
{\bar{\bar{t_i}}}  <  {\bar{\bar{t}}} < {\bar{t_1}}   \\
\bar{a_{II}}(\bar{t}) & = & \bar{A_{II}} {(\bar{t}-\bar{t_{II}})}^R 
 \ \ \ \ \ \ \ \   \bar{t_1} < \bar{t}  < \bar{t_2}\nonumber \\
a_{III}(t) & = & A_{III} {(t)}^M 
\ \ \ \ \ \ \  \ \ \ \ \bar{t_2} <  t < {t_0} \nonumber
\ees
with continuous transitions at $\bar{t_1}$ and $\bar{t_2}$ for both the 
scale factor
and first derivatives with respect to the descriptive cosmic time-type 
variables $\bar{\bar{t}}$, $\bar{t}$ and $t$.
In terms of the standard observational times $t_r$ and $t_m$, the 
transitions $\bar{t_1}$, $\bar{t_2}$
and the beginning of the inflationary stage description $\bar{\bar{t_i}}$ are
expressed as:
\bes  
\bar{t_1} & = & {R \over{M}} t_r + \left(1 - {R \over{M}}\right) t_m 
\ \ \ \ \ \ , \ \ \ \ 
\bar{t_2}  =  t_m \nonumber \\
\bar{\bar{t_i}} & = &
 \left({R \over{M}} + {Q\over{M}}{{t_r-t_i}\over{t_r-t_I}} \right)t_r
+ \left(1 - {R \over{M}}\right) t_m \label{Ti}
\ees
and the set of parameters of the scale factor (\ref{Descr})
can be written too in terms of
$t_r$, $t_m$ and the global scale factor $\bar{A_{II}}$:
\bes 
\bar{\bar{t_I}} & = & {t_r} {\left({R\over{M}}+{Q\over{M}}\right)} + {t_m}
{\left({1 - {R \over{M}}}\right)} \ \ \ , \ \ \  
\bar{t_{II}}  =  \left({1-{R\over{M}}}\right) t_m \nonumber \\
\bar{\bar{A_I}} & = & \bar{A_{II}} {\left({Q \over{M}}\right)}^Q 
{\left({R \over{M}}\right)}^R {t_r}^{R+Q} \ \ \ \ \ \ \ , \ \ \  \
A_{III}  =  \bar{A_{II}} {\left({R \over{M}}\right)}^{R} {t_m}^{R-M} 
\ees

Notice that the intermediate descriptive scale factor could not 
be physical, but this is
not important under the purposes here. Since the
description is linked with observational information and it satisfies
the proper scale factor ratii, our computation will be equivalent to
that made on a full and physical model. These are equivalents unless 
the features left on gravitational wave production by the details of an
 in any case unknown dynamics of transition and minimized through
their modelization in satisfactory continuos way, as made here.
The time variables $\bar{\bar{t}}$ of inflationary stage,
and ${\bar{t}}$ of radiation stage are not a priori exactly equal to 
the physical time coordinate at rest frame (multiplied by $c$), but
transformations (dilatation plus translation) of it. The low energy 
effective action equations from where the scale factor, dilaton and
equation of state
have been extracted, allows this transformations.
With this treatment of cosmological scale factor, we will attain 
computations free of "by hand" parameters, and with full predictibility
as can be seen in the next sections. 

It is also useful to express this model in conformal time variables $\eta$,
defined as $d \eta = \frac{d t}{a(t)}$. We will have a conformal time 
variable for each
stage ($\bar{\bar{\eta}}$, $\bar{\eta}$ and $\eta$) oughted to the different 
scale factor shape and different descriptive temporal variable on
which we integrate. In this way, the conformal
time variable constructed enjoys continuity on their derivatives
along the transitions
at $\bar{t_1}$ and $\bar{t_2}$ \cite{p1}.
\besf
\left. {\frac{d{\bar{\bar{\eta}}}}{d{\bar{\bar{t}}}}}\right|_{{{\bar{t_1}}}^-}
 = \left. {\frac{d{\bar{\eta}}}{d{\bar{t}}}}\right|_{{{\bar{t_1}}}^+} 
\ \ \ \ \ , \ \ \ \ \ 
\left. {\frac{d{\bar{\eta}}}{d{\bar{t}}}} \right|_{{{\bar{t_2}}}^-}  =  
\left. {\frac{d{\eta}}{d{t}}} \right|_{{{\bar{t_2}}}^+} 
\eesf
We can impose also continuity conditions in order to have
a conformal time uniquely defined at transitions. 
$ \eta_1  =  \bar{\eta}({\bar{t_1}}) \equiv  \bar{\bar{\eta}}
({\bar{t_1}}) $ and $ \eta_2  =  \eta{({\bar{t_2}})} \equiv 
 \bar{\eta}(\bar{t_2})$
(In this respect, different ways can be found in literature, 
see \cite{p1}). We use the remaining freedom in put a 
simplicity condition, asking a simple conformal time
dependence in radiation dominated stage.
It yields the
next expressions for the scale factor in conformal time: 
\bes \label{EDescr} 
{{a_I}}(\bar{\bar{\eta}}) & = & {\alpha_I} {(\eta_I - {\bar{\bar
{\eta}}})}^{-q} \ \ \ \ \ \ \ \ \ \ \eta_i < \bar{\bar{\eta}} < \eta_1 \\
{a_{II}}(\bar{\eta}) & = & {\alpha_{II}} {(\bar{\eta})}^r 
\ \ \ \ \ \ \ \ \ \ \ \ \ \ \ \ \ \eta_1 < \bar{\eta} < \eta_2  \nonumber  \\
a_{III}(\eta) & = & {\alpha_{III}} {(\eta_{III} + \eta)}^m \ \ \ \ \  
\eta_2 < \eta  \nonumber
\ees
with continuous and suddenly transitions (continuity of scale factor and
first derivative with respect to conformal time) at $\eta_1$ and $\eta_2$.
The six parameters satisfy these matching relations:
\bes
\alpha_I =  \alpha_{II} {\left(\frac{q}{r}\right)}^q {(\eta_1 
)}^{r+q} \ \ \ \ \ & , &\ \ \ \ 
\alpha_{III}  =  \alpha_{II} {\left(\frac{m}{r}\right)}^{-m} 
{(\eta_2)}^{r-m} \label{MAT1}\\
\eta_I  =  \eta_1 \left(1+\frac{q}{r}\right)
\ \ \ \ \ \ \ \ \ \ \ \ & , & \ \ \ \ \ 
\eta_{III} = \eta_2 {\left(\frac{m}{r}-1\right)} \nonumber
\ees
Because its construction on a cosmic time description with fully 
continuity conditions,  all the parameters involved in this conformal
time scale factor have a complete and unique expression as 
functions of the observational standard transition times $t_r$ and $t_m$,
beginning of inflationary stage $t_i$, the exponents $Q$, $R$,$M$
of cosmic time dependences  
and the global scale factor constant $\bar{A_{II}}$:
\bes
q & = & {Q\over{Q+1}} \ \ \ , \ \ \ r = {R \over{1-R}} 
\ \ \ , \ \ \ m = {M \over{1-M}} \label{SET1} \\
\eta_I & = & {{R+Q}\over{R(Q+1)(1-R)}} {{{{({R\over M}t_r)}^{1-R}}}\over
{\bar{A_{II}}}} \nonumber \\
\alpha_I & = & {\left(\bar{A_{II}}{\left({R\over {M}}\right)}^R 
{\left({Q\over {M}}\right)}^Q
{1 \over{{(Q+1)}^Q}} \; {t_r}^{R+Q} \right)}^{1\over{Q+1}} \nonumber \\
\alpha_{II} & = & {\bar{A_{II}}}^{1\over{1-R}} {(1-R)}^{R\over{1-R}} 
\nonumber  \\
\eta_{III} & = & {{M-R}\over{R(1-M)(1-R)}} 
{{{({R\over M} t_m)}^{1-R}}\over{\bar{A_{II}}}} \nonumber  \\
\alpha_{III} & = & {\left(\bar{A_{II}} {(1-M)}^M {\left({R\over M}\right)}^R
{t_m}^{R-M}\right)}^{1\over{1-M}} \nonumber 
\ees
Meanwhile, the conformal time transitions $\eta_1$, $\eta_2$ and $\eta_i$
have the expressions:
\bes
\eta_1 & = & \frac{{\left(\frac{R}{M}\:t_r\right)}^{1-R}}{(1-R){\bar{A_{II}}}} 
\ \ \ \ \ , \ \ \ \ \ \ 
\eta_2  =  \frac{{\left(\frac{R}{M}\:t_m\right)}^{1-R}}{(1-R){\bar{A_{II}}}} 
\label{et1ob} \\
\eta_i & = & \frac{{\left(\frac{R}{M}\:t_r\right)}^{1-R}}{(Q+1){\bar{A_{II}}}}
{\left[\frac{R+Q}{R(1-R)} \: - \: \frac{Q}{R}{\left(\frac{t_i-t_I}
{t_r-t_I}\right)}^{Q+1}\right]} \nonumber
\ees
It is also possible give expressions for the conformal time in each
stage as function of the proper cosmic time:
\bes 
\bar{\bar{\eta}} & = & {{{({R\over M} t_r)}^{1-R}}\over{\bar{A_{II}}
(Q+1)}} \left[{{R+Q}\over{R(1-R)}}-{Q\over R} 
{\left({{t_I -t}\over{t_I-t_r}}\right)}^{Q+1}\right] \label{etcont1} 
\ \ \ t \in (t_i, t_r) \\
\bar{\eta} & = & {{{({R\over M} t)}^{1-R}}\over{\bar{A_{II}}(1-R)}} 
\ \ \ \ \ \ \ \ \ \ \ \ \  \ \ \ \ \ \ \ \ \ \ \ \ \ \ \ \  \ \ \ \ \ \ 
\  t \in (t_r, t_m) \nonumber \\
\eta & = & {{{({R\over M} t_m)}^{1-R}}\over{\bar{A_{II}}(1-M)}}
\left[{{R-M}\over{R(1-R)}} + {M\over R} {\left({t\over{t_m}}\right)}^{1-M}
\right] \ \ \ \ t \in (t_m, t_0) \nonumber
\ees

Finally, it is useful for further work to compute here the value for the
conformal time 
and scale factor at current time
$\eta_0$. If we define $t_0 = c {\mathcal{T}}_0$ and from eqs.(\ref{etcont1}),
it is easily seen:
\bes
{\eta}_0 & = & {{{({R\over M} t_m)}^{1-R}}\over{\bar{A_{II}}(1-M)}}
\left[{{R-M}\over{R(1-R)}} + {M\over R} 
{\left({{t_0}\over{t_m}}\right)}^{1-M}\right] \label{ethoy}
\ees
The scale factor written in conformal time for current time
takes the form $a_{III}({\eta}_0)  =  {\alpha_{III}} 
{(\eta_{III} + {\eta}_0)}^m$. From eqs.(\ref{SET1}) and (\ref{ethoy})
we translate this value 
to observational transition values:
\bes \label{Ahoy}
a_{III}({\eta}_0) & = & {\bar{A_{II}}} {\left(\frac{R}{M}\right)}^R
\; {t_0}^M\; {t_m}^{R-M}
\ees

The last point is to make an approach for the dilaton field. This is
considered practically constant from the beginning of radiation
dominated stage until the current time. Their value can be supposed
coincident with the value at exit inflation time in a sudden but
not continuous transition for the dilaton, since no one of 
their temporal derivatives can match this asymptotic behaviours.
\be \label{dilr}
{\phi}_{II} \: = \: {\phi} (\eta_1) \: = \: \phi_1
\ee
Remember the expression for dilaton in inflation dominated
stage, that gives
\be \label{dili}
{\phi}_{II} \: = \: {\phi}_I + 2 d \ln a(\eta_1)
\ee

\begin{table}[h]
\centering
\begin{tabular}{||c|c||c|c||}
\hline
\hline
\multicolumn{2}{||c||} {\it temporal dependence} 
& \multicolumn{2}{c||}{{\emph{String Driven Cosm.B.}}} \\
\multicolumn{2}{||c||} {\it parameters} & $\: d \:$ & $d=3$ \\ \hline
{\it Cosmic} & infl.  {\bf Q} & $\frac{2}{d+1} $ &$\  $$\frac{1}{2}$ $\ $ \\
\cline{2-4}
{\it Time} & rad.  {\bf R} & $\frac{2}{d+1} $ & $\frac{1}{2}$ \\
\cline{2-4}
{\it description} & mat.  {\bf M} & $\frac{2}{d} $ & $\frac{2}{3}$ \\
\hline
{\it Conformal}& infl. {\bf{q}} & ${2\over{d+3}}$ & ${1\over3}$ \\
\cline{2-4}
{\it Time} & rad. {\bf{r}} & ${2\over{d-1}}$ & $1$ \\
\cline{2-4}
{\it description} & mat. {\bf{m}} & ${\frac{2}{d-2}}$ & $2$ \\
\hline \hline
\end{tabular}
\caption{\label{t.cos} Temporal Dependences in the String Driven 
Cosmological Backgrounds}
\vspace{5pt}
{\parbox{135mm}{\footnotesize Values of the exponents in the temporal 
dependences for the inflationary, radiation dominated and matter dominated 
stages in String Driven Cosmological Background. The parameters 
$Q$, $R$ and $M$ appear in the cosmic time description, whereas the 
parameters $q$, $r$ and $m$ are in the conformal time one. 
Particularization to
the three dimensional case is given.}}
\end{table}

\section{Generation of Gravitational Wave Perturbations in String Cosmology}
\setcounter{equation}0

{\margen} After obtaining the minimal cosmological model for the scale factor
in the framework of string cosmology, we will compute one of its
important observational consequences, the stochastic background of 
relic gravitational waves. As first noticed by Grishchuk \cite{gkprd}, relic 
gravitational waves are inevitably generated in practically all cosmological 
models oughted to their variable gravitational field.

The generation of gravitational wave backgrounds can be studied classically 
as  propagation and amplification by 
the expanding background of one initial spectrum of tensorial perturbations 
on the metric. From the quantum mechanical point of view, a particle 
production mechanism occurs and gravitational waves of every wavelenght 
are generated as gravitons from an initial vacuum state. 
This translates into a combined 
classical-quantum mechanical formalism. (See for instance \cite{abha}
and \cite{gksqz}). Gravitons produced at scales near the Planck time 
must exist today in the form of gravitational waves of wavelenghts to be 
determined by the propagation equation. \cite{gkz}.

The number or particles produced 
between an initial -asymptotic- vacuum state and a final state, where particles
-gravitons- has been created by the effect of the changing gravitational field 
can be obtained by computing the Bogoliubov coefficients \cite{A}. 
Grishchuk \cite{gkprd} has a similar approach modelizing the 
changing gravitational  field as a potential barrier. 
Each polarization degree of freedom of 
tensorial perturbation satisfies the same propagation equation that a
massless scalar field (\cite{fpa},\cite{gks},\cite{wh}). But in frameworks
with a dilaton field, attention must be
put in accounting for the dilaton contribution to the gravitational field 
change and to the 
metric perturbation, breaking the validity of reported treatments \cite{gdil}. 

In the next, we summarize and follow the treatment of tensorial perturbations
in cosmological backgrounds within the Brans-Dicke frame developped 
in ref.\cite{gdil}. That is necessary for accounting of
the full effect of the dilaton field and for the consistency with the framework
in which the string cosmology solutions have been obtained.
We will show that this formalism includes as a particular case 
(no dilaton field) the formalism developped by Allen 
\cite{A} and Grishchuk \cite{gkprd} in the framework of quantum field theory.

The general action written in Brans-Dicke frame takes the form:
\be
S = - \frac{c^3}{16 \pi G_D} \int d^{d+1} x \sqrt{\mid g \mid} e^{- \phi}
\left(R - \omega \: g^{\mu \nu} \: \partial_{\mu} \phi \: \partial_{\nu} \phi 
\right) + S_M
\ee
where $S_M$ represents the contribution of matter sources such that its
stress tensor is $\sqrt{\mid g \mid} T_{\mu \nu} = 2 \frac{\delta S_M}
{\delta g^{\mu \nu}}$, antisymmetric tensors and dilaton potential are 
vanishing and $\omega$ is the usual Brans-Dicke parameter. 
For $\omega = \infty$ General Relativity expressions are found and for 
$\omega = -1$ this action coincides with the low energy effective string 
action eq.(\ref{action}).
By varying with respect to the metric $g_{\mu \nu}$ 
and to dilaton $\phi$ the following two equations are obtained:
\bes 
{R_{\mu}}^{\nu} + \nabla_{\mu} \nabla^{\nu} \phi + (\omega + 1) [{\delta_{\mu}}
^{\nu}({(\nabla \phi)}^2 - \Box \phi) - {\nabla}_{\mu} \phi {\nabla}^{\nu} 
\phi] &  = & \frac{8 \pi G_D}{c^4} e^{\phi} {T_{\mu}}^{\nu} \ \ \ \ \ \  
\label{firsteqbd1}\\
R + \omega {(\nabla \phi)}^2 - 2 \omega \ \Box \phi & = & 0 \label{firsteqbd2}
\ees
where $\Box={\xi}^{\mu \nu}\nabla_{\mu}\nabla_{\nu}$ is the covariant 
D'Alambertian operator. Again, 
the case $\omega=-1$ recoveres the low energy string effective equations 
(\ref{leeeq}).

The free-linearized wave equation for a generic metric fluctuation
$h_{\mu \nu} = \delta g_{\mu \nu}$ is
obtained by expressing the perturbed metric as
${\hat{g}}_{\mu \nu} = g_{\mu \nu} + \delta g_{\mu\nu}$, 
and all the sources are considered fixed
$\delta {T_{\mu}}^{\nu} = \delta \phi  = 0$,
second order (and higher) terms in $h_{\mu \nu}$ are neglected. 

In the following, the study 
is made only for the tensorial 
component of the metric fluctuation $h_{\mu \nu}$. 
The general case for the metric is considered: 
an anisotropic background, with $d$ 
expanding external spatial dimensions with scale factor $a(t)$ and $n$ 
contracting internal dimensions with scale factor $b(t)$,
both flat and Euclidean. Expressed in the
synchronous gauge, the metric is:
\be \label{metn}
g_{\mu \nu} = (1, -a^2(t)\gamma_{i j}(x), -b^2(t) \gamma_{a b}(y))
\ee 
where again the component (0,0) is written in lenght units and corresponds to 
the variable $t = c \mathcal{T}$, indexes $i, j$ run from $(1 \dots d)$ and 
indexes $a,b$ run from $(d+1 \dots d+n+1)$. 

The metric perturbation is considered propagating 
only in the external spatial dimensions, decoupled from the sources and 
purely tensorial:
\bes
h_{\mu \nu} & = & h_{\mu \nu}({\bf x}, t) \\
h_{0 \mu} & = & h_{a \mu} \ \ = \ \ 0 
\ees
That means, a pure gravitational wave that can
be expressed in the transverse-traceless gauge, thus 
$g^{\mu \nu} \ h_{\mu \nu}  =  0$ and 
$\nabla_{\nu} {h_{\mu}}^{\nu}  =  0$.
With these conditions, eq.(\ref{firsteqbd2}) is trivially satisfied and 
from (\ref{firsteqbd1}) the free-linearized wave equation for the 
tensorial metric perturbation $h_{\mu \nu}$ is obtained:
\be \label{eqlinn}
\delta {R_{\mu}}^{\nu} + {1\over{2}} \; \dot{\phi} \; \dot{h}_{\mu \alpha} 
\;g^{\nu \alpha} - h^{\nu \alpha} \nabla_{\mu} \nabla_{\alpha} \phi = 0
\ee
It must be noticed that this equation is $\omega$-independent and therefore
it is valid in general for all Brans-Dicke cases and in particular for the 
low energy effective string action. Eq.(\ref{eqlinn}) takes a shorter form 
when the whole information about the metric is introduced \cite{gdil}.
\be \label{short} 
\Box {h_i}^j - \dot{\phi} \dot{{h_i}^j} = 0
\ee

Writting the perturbation in terms of eigenstates of the Laplace 
operator of wavenumber $k$: 
$\nabla^2 {h_i}^j(k) = - k^2 {h_i}^j(k)$, 
eq. (\ref{short}) takes the form: 
\be \label{eqtc}
\ddot{{h_i}^j} + (dH + nF - \dot{\phi}) \dot{{h_i}^j} + 
{\left({k\over{a}}\right)}^2 {h_i}^j = 0
\ee
here dot means derivative with respect to $t$ (remember 
$t=c \mathcal{T}$),
$H=\frac{\dot{a}}{a}$ and $F=\frac{\dot{b}}{b}$.
The most suitable form of eq.(\ref{eqtc}) is obtained by writting it in 
conformal time $d\eta = \frac{dt}{a(t)}$. It is
convenient to define a variable $\psi$:
\be \label{psi}
{\psi_i}^j = {h_i}^j \; a^{{d-1}\over{2}} \; b^{n\over{2}} \; 
e^{-{{\phi}\over{2}}} \ \ \ \ ,
\ee
in such a way that ${\psi_i}^j$ is identified with each one of the two 
polarization modes of the perturbation $h_{i j}$. From eq.(\ref{eqtc})
it is immediately seen that each polarization mode $\psi$ satisfies the
corresponding equation written in conformal time:
\bes \label{EQ}
\psi^{''} +(k^2 - V(\eta))\psi = 0 & &
\ees
where :
\bes \label{POT}
V(\eta) & = & {{d-1}\over{2}} {{a^{''}}\over{a}} + {n\over{2}}
{{b^{''}}\over{b}} - {{{\phi}^{''}}\over{2}} + {1\over{4}}(d-1)(d-3)
{\left({{a^{'}}\over{a}}\right)}^2 +
{1\over{4}}n(n-2){\left({{b^{'}}\over{b}}\right)}^2 \nonumber \\
& + & {1\over{4}}{{\phi}^{'}}^2 + {1\over2}n(d-1){{a^{'} b^{'}}\over{a \; b}} 
- {1\over2}(d-1){\left({a^{'}}\over{a}\right)}\phi^{'} - 
{n\over2}{\left({b^{'}}\over{b}\right)}\phi^{'}
\ees

Thus, the problem of metric fluctuation propagation is reduced to a
second order differential Schr\"{o}dinger type equation, where the 
``potential'' $V(\eta)$ takes into account the conformal time derivatives 
of the background as well as those of dilaton fields; that means, the whole
variable gravitational field ``pumping'' the generation and amplification of 
gravitons.\cite{gdil}

Following our definition of the variable $t=c\mathcal{T}$ in length units, 
the conformal time variable $\eta$ has also length units. Thus, the 
``potential'' (\ref{POT}) have units $({\mathrm{length}})^{(-2)}$ and the 
wavenumber
$k$ has units $({\mathrm{length}})^{(-1)}$.

As it was pointed out in \cite{gdil}, this treatment generalizes to higher
dimensional backgrounds the treatment more widely work, usually based on 
the pioneering works of Grishchuck \cite{gkz},
\cite{gks}. It generalizes also other early works like those of Allen \cite{A}.

\pagebreak[4]

\subsection{Particular Cases}

{\bf{String Driven Case}}

\margen In order to compute perturbations in our cosmological background, 
we will consider the suitable particularization of eqs.(\ref{EQ}) and 
(\ref{POT}). For arbitrary $d$ and $n=0$ 
eqs.(\ref{EQ}) and (\ref{POT}) take the form:
\bes{\label{eq:SD}}
\psi^{''} & + & (k^2 - V(\eta))\psi = 0 \\
V(\eta) & = & {{d-1}\over{2}} {{a^{''}}\over{a}} - {{{\phi}^{''}}\over{2}} +
{1\over{4}}(d-1)(d-3){\left({{a^{'}}\over{a}}\right)}^2  \nonumber \\
& + & {1\over{4}}{{\phi}^{'}}^2 -
{1\over{2}}(d-1){\left({{a^{'}}\over{a}}\right)}{\phi}^{'} \label{V:SD}
\ees
The perturbation variable $\psi$ is written as:
\bes \label{psi:SD}
{\psi_i}^j & = & {h_i}^j a^{{d-1}\over2} e^{-{{\phi}\over2}} \ \ \ \ ,
\ees
and as consequence, $\psi_0^{\mu} = \psi_{\mu}^0 = 0$.
Equivalently, the tensorial perturbation is:
\bes \label{hm:SD}
h_{i j} & = & {{\psi}_i}^j a^{-\frac{d+3}{2}} e^{\frac{\phi}{2}} \\
h_{0 \;\mu} & = & h_{\mu \; 0} \ = \ 0  \nonumber \\
h_{a \; \mu} & = & h_{\mu \; a} \ = \ 0 \nonumber
\ees
where the index $\mu$ runs now from $1 \dots d$.

\noindent {\bf No Dilaton Case}

It is possible to extract from eqs.(\ref{eq:SD}), (\ref{V:SD}) and
(\ref{psi:SD}) the case without dilaton field, in which we obtain the 
set of equations:
\bes{\label{eq:A}}
\psi^{''} & + & (k^2 - V(\eta))\psi = 0 \\
V(\eta) & = & {{d-1}\over{2}} {\left({a^{''}}\over{a}\right)} +
{1\over{4}}(d-1)(d-3){\left({{a^{'}}\over{a}}\right)}^2 \label{V:A}
\ees
The perturbation variable $\psi$ is:
\bes \label{psi:A}
{\psi_i}^j & = & {h_i}^j \; a^{{d-1}\over2}
\ees
and the metric tensorial perturbation is:
\besf
h_{i\;j} & = & {{\psi}_i}^j \; a^{-\frac{d+3}{2}} \\
h_{0\;\mu} & = & h_{\mu\;0} \; = \; 0 
\eesf

The case of three spatial dimensions gives the simplest 
expressions:
\bes{\label{eq:3}}
\psi^{''} & + & (k^2 - V(\eta))\psi = 0 \\
V(\eta) & = &  {\left({a^{''}}\over{a}\right)} \label{V:3}
\ees
where the perturbation is simply written
\bes \label{psi:3}
{\psi_i}^j & = & a \; {h_i}^j 
\ees
with $h_{i\;j}  =  {{\psi}_i}^j \; a^{-3}$ and 
$h_{0\;\mu}  =  h_{\mu\;0} \; = \; 0 $. 
This particular (3+1)-dimensional case gives the same set of 
equations ((\ref{eq:3})-(\ref{psi:3})) that can be
found in Grischuhck treatments \cite{gks}. These equations 
can be also identified with the formalism found in \cite{A}.

For the sake of completeness, we compare now with the 
linearized Einstein equations for an
homogeneus, isotropic and spatially flat metric $g_{\mu \nu}$
which in the conformal time gauge is:
\be{\label{eq:metrica}}
ds^2 = {a(\eta)}^2 ({-d\eta}^2 + {d{\bf x}}^2)
\ee
\indent A gravitational perturbation $h_{\mu \nu}=a^2(\eta) f_{\mu \nu}$ 
of comoving wave number ${\bf k}$ is overimposed such that:
\be 
{\hat{g}}_{\mu \nu} = a^2(\eta) (\eta_{\mu \nu} + f_{\mu \nu})
\ee

By performing a Fourier expansion, 
$f_{\mu \nu}$ can be described as:
\be{\label{eq:pert}}
f_{\mu \nu} (\eta, {\bf x}) = \varphi(\eta) e_{\mu \nu} ({\bf k}) \exp 
(\imath{\bf k x})
\ee
where $e_{\mu \nu}$ is the constant polarization vector. 
It can be shown that the amplitude perturbation satisfies the
equation \cite{A}:
\be \label{ela}
\varphi''(\eta) + (d-1) {{a'(\eta)} \over {a(\eta)}} \varphi' (\eta) + 
k^2 \varphi = 0
\ee
where $k=\mid{\bf{k}}\mid$, {\bf{k}} is the comoving wave number and prime
stands for conformal temporal derivatives. This equation is equivalent to 
eq.(\ref{eqtc}) in the Brans-Dicke formalism, provided particularization
to isotropic cases ($F=0$) and no dilaton dynamics ($\dot{\phi}=0$).
For $d=3$, eq.(\ref{ela}) is the known equation found in the 
literature (\cite{A}).

Let us write:
\bef
\varphi(\eta) = {a(\eta)}^{-\left(\frac{d-1}{2}\right)} v(\eta) \ \ \ .
\eef
Then, we have
\be{\label{eq:ampu}}
u''(\eta) + {\left[{{d-1}\over2}{\left({{3-d}\over2}{\left({a'(\eta)\over
{a(\eta)}}\right)}^2 - {{a''(\eta)}\over{a(\eta)}}\right)} + k^2\right]} 
u(\eta) = 0
\ee
that, in the particular three spatial dimensions take the form:
\besf
\varphi(\eta) & = & \frac{u(\eta)}{a(\eta)} \\
v'(\eta) & = &-{{a'(\eta)}\over{a(\eta)}} v(\eta) \\
u''(\eta) & + & {(k^2 - V(\eta))}u(\eta) = 0 \ \ \ \ \ \ \ \
V(\eta) = {{a''(\eta)}\over{a(\eta)}} 
\eesf
This shows the equivalence with the Grishchuk formalism
and with the (3+1)-dimensional particularization of
the No Dilaton Case showed above.

\subsection{The Power Spectrum}

\margen We perform an expansion 
of the vacuum state (amplitude of perturbations at the classical level) in 
positive- and negative- frequency modes in each stage. The Bogoliubov 
coefficients mix the modes before and after the
transition or the graviton
creation-annihilation operators $a_k$ and ${a_k}^+$. 
$\mid\beta_k\mid^2$ gives the number of particles of 
comoving frequency $k$ generated on the vacuum state.
This is the ``initial'' state (at a time sufficiently 
early before the transition to radiation dominated stage), so
the corresponding Bogoliubov coefficient in the 
radiation dominated stage yields the total number of particles created
during the whole inflationary era with respect that initial state.

We will determine the Bogoliubov coefficients $\alpha(k)$ and 
$\beta(k)$ for every wavenumber perturbation $k$ by considering a continous
matching of the perturbation amplitude at the transition time $\eta_1$ 
between inflation and radiation dominated stages. Thus, we will impose
continuity of the amplitude perturbation -treated as quantum state- as well 
as of its first conformal time derivative at the sudden transition at
conformal time $\eta_1$. 
This is similar to work with the picture of a potential barrier generated 
by the variable gravitational field {\cite{gks}}. Gravitational waves
would interact with this barrier. By describing one state 
``out'' the barrier, one state ``in'' with frozen amplitude and a comeback 
to ``out'', the Bogoliubov coefficients can be computed.
But as above said, the contribution of the dilaton change
can be missed.

The stochastic background of gravitational 
waves generated in the 
cosmological  transitions is described by the power spectrum 
$P(\omega)d\omega$, which is the energy density
contained in gravitational waves with frequencies in the interval between
$\omega$ and $\omega + d\omega$; $\omega$ is the proper frequency
at the detection time related to the comoving wavenumber $k$ through:
\be \label{deom}
\omega = \frac{c\;k}{a(\eta_0)}
\ee
The total energy density in gravitational waves is computed as 
$\rho_{GW} \; = \; \int P(\omega) d\omega$ \cite{A}.
The power spectrum  $P(\omega) d\omega$ is given by 
$ P(\omega) d\omega=
2 \hbar \omega {\mid\beta\mid}^2 d\omega dN$ where $dN$ is
the density of states, which in three spatial dimensions (as we have at the
detection moment) is 
$dN={{\omega^2}\over{2 {\pi}^2 c^3}}d\omega$;
${\mid\beta\mid}^2$ is the number of gravitons created in each frequency 
interval and each polarization state. This comes multiplied by the two 
polarization degrees of freedom of the tensorial metric perturbation 
and by the energy  $\hbar \omega$ of each one:
\be{\label{eq:pomega}}
P(\omega)d\omega=  {{\hbar \omega^3}\over{\pi^2 c^3}}
{\mid\beta\mid}^2 d\omega 
\ee
Thus, the power spectrum $P(\omega)$ has dimensions 
$[P(\omega)]= \frac{{\mathrm{energy}}\;{\mathrm{time}}}{\mathrm{volume}} 
\; = \; \frac{{\mathrm{mass}}}{{\mathrm{time}}\;{\mathrm{lenght}}}$.

It is usual to write the power spectrum as fraction of the critical 
energy density by octave, which is the
closest form to gravitational wave detection procedures:
\bef
\Omega_{GW} \: = \: \frac{1}{\rho_c} \; \frac{d \rho_{GW}(\omega)}
{d \ln \omega}
\eef
here $\rho_{GW}$ is the energy density contained in gravitational waves 
and $\rho_c$ is the critical energy density 
$\rho_c = {{3 \; {H_0}^2}\over{8 \pi G}}$.
This is equivalent to compute
\be \label{ome}
\Omega_{GW} \: = \: \frac{\omega}{\rho_c}\; P(\omega)
\ee
Obviously, $\Omega_{GW}$ is dimensionless
$[\Omega_{GW}] \; = \; 1$.

\section{Power Spectrum in String Driven Cosmology}
\setcounter{equation}0

\subsection{No Dilaton Case}

\margen We begin our computations of power spectrum of gravitational waves 
with the 
simplest case:
we will neglect in this the whole dilaton role. Thus, we will solve
eqs.(\ref{eq:A}), (\ref{V:A}) and the reduced amplitud
perturbation given by (\ref{psi:A}).

In the inflationary stage, from eqs.(\ref{EDescr}) and (\ref{V:A}),
the "potential" $V(\eta)$ is given by:
\bes \label{V:ndi}
V(\eta) & = & \left({\left({{d-1}\over{2}}q \: + \frac{1}{2}\right)}^2 - 
{1\over{4}}\right) {(\eta_I-\bar{\bar{\eta}})}^{-2}
\ees
The general solution of eq.(\ref{eq:A}) with the potential
(\ref{V:ndi}) is given in terms of Bessel functions:
\be \label{vacgen}
\psi = {(\eta_I-{\bar{\bar{\eta}}})}^{\frac{1}{2}} {\left(
 C_1 {{\mathcal{H}}_{\nu}}^{(1)}(k(\eta_I-{\bar{\bar{\eta}}}))
+ C_2 {{\mathcal{H}}_{\nu}}^{(2)}(k(\eta_I-{\bar{\bar{\eta}}})) \right)}
\ee 
with index $\nu$:
\be \label{nu:A}
\nu = \pm \frac{1}{2}{\left((d-1)q + 1\right)}
\ee
Or, from eq.(\ref{Sdrin}) in terms of the number $d$ of the
spatial dimensions.
\be \label{nu:ND}
\nu =  \frac{3d+1}{2(d+3)}
\ee
It gives  $\nu =  \frac{5}{6}$ for three spatial dimensions. 

The coefficients $C_1$ and $C_2$ are fixed
by the choice of initial or boundary conditions, i.e. by 
the inflationary vacuum state \cite{A}. 
Thus, the asymptotic expression of (\ref{vacgen}) for early times 
should contain only positive 
frequency modes. 

The argument of the Hankel functions involved can be expressed in terms
of the cosmic time. From eq.(\ref{MAT1}):
\be
z \: \equiv \: k(\eta_I-{\bar{\bar{\eta}}}) \: = \: 
k \left(\eta_1(1+\frac{q}{r})-
{\bar{\bar{\eta}}}\right) \ \ ,
\ee
and with the expressions relating the conformal time description with the
observational transition times and cosmic time 
eqs.(\ref{SET1}), (\ref{et1ob}) and (\ref{etcont1}), we get:
\be
z \: = \: k \; {\left(\frac{R}{M}t_r\right)}^{1-R} \frac{Q}{R(Q+1)}
\frac{1}{\bar{A}_{II}} {\left[{\left(\frac{t_I-t}{t_I-t_r}\right)}^{Q+1}
\right]}
\ee
As discussed in section II
(see eq.(\ref{sdrin}) and comments below), the parameter $t_I$ is 
always greater than the exit of inflationary stage $t_r$, thus the
argument $z$ is real and $z > 0$ for all $t < t_r$. For time very early
in the inflationary stage $t \rightarrow t_i$ and $t << t_r$, the
argument $z$ approaches its maximum value. Thus, we must identify
the large argument regime with the vacuum state at enough early 
inflationary stage (usually refered as well below the transition to
radiation dominated epoch). 

The asymptotic behaviours of Hankel functions {\cite{taa}}:
and the positive frequency mode condition at early stages, fix
$C_2 = 0$. The remaining freedom in the coefficient $C_1$ is 
linked to the normalization of the vacuum inflationary state 
and we include it in the normalization
coefficient ${\mathcal{N}}$.  In this way, the reduced amplitude 
perturbation
in the inflationary stage (before the transition) is written as:
\be \label{vacinf}
{\psi}_I \: = \: {\mathcal{N}} {\left(\eta_I-
\bar{\bar{\eta}}\right)}^{\frac{1}{2}}
{{\mathcal{H}}_{\nu}}^{(1)}\left(k(\eta_I-\bar{\bar{\eta}})\right)
\ee

In the radiation dominated stage, from the scale factor 
eq.(\ref{EDescr}), the potential 
(\ref{V:A}) takes the form:
\bes \label{V:ndr}
V(\eta) & = & \left({\left({{d-1}\over{2}}r \: - \frac{1}{2}\right)}^2 - 
{1\over{4}}\right) {\bar{\eta}}^{-2}
\ees

And the general solution of eq.(\ref{eq:A}), with eq.(\ref{V:ndr}) is given 
again in terms of  Hankel functions:
\be \label{radgen}
\psi = {\bar{\eta}}^{\frac{1}{2}} \; {\left(
  {\mathcal{D}}_1 {{\mathcal{H}}_{\mu}}^{(1)}(k{\bar{\eta}})
+ {\mathcal{D}}_2 {{\mathcal{H}}_{\mu}}^{(2)}(k{\bar{\eta}}) \right)} \ \ ,
\ee 
but now with index $\mu$:
\be \label{A:mu}
\mu = \pm \frac{1}{2}{\left((d-1)r - 1\right)} \ \ \ .
\ee
Since the radiation dominated stage has
$r \; = \; \frac{2}{d-1}$ (eq.(\ref{Sdrad})), it makes: 
\be
\mu = \pm \; \frac{1}{2}
\ee
In this case, by using the properties of Hankel functions with
fractional index (See \cite{taa}) , $\psi$ eq.(\ref{radgen}) 
has the expression:
\bes
\psi  & = & - {i} \: \sqrt{\frac{2}{\pi k}} {\left(
  {\mathcal{D}}_1 e^{ik\bar{\eta}}
- {\mathcal{D}}_2 e^{-ik\bar{\eta}} \right)}
\ees
which can be written also in the way
\be \label{vacrad}
{\psi}_{II}  =  i \: {\mathcal{M}} \sqrt{\frac{2}{\pi k}}
{\left(\alpha e^{-i k(\bar{\eta} - \eta_1)} + \beta e^{i k(\bar{\eta}-\eta_1)}
\right)} \ \ \ \ \ .
\ee

Here we have defined ${\mathcal{M}}$ a suitable normalization
constant and the coefficients 
$ \alpha  =  \frac{{\mathcal{D}}_2}{\mathcal{M}} \;  e^{-\;ik{\eta}_1}$
and 
$\beta  = - \frac{{\mathcal{D}}_1}{\mathcal{M}} \; e^{ik{\eta}_1}$
are the Bogoliubov's coefficients, since they are
multiplying the positive- and negative-frequency pure modes.

With the expressions for the reduced amplitude perturbation
in each stage (\ref{vacinf}) and (\ref{vacrad}), where we remember the
parameter $\nu$ has the value given by (\ref{nu:A}), 
we will determine the value of Bogoliubov's coefficients $\alpha$ and $\beta$ 
by matching the
reduced amplitude perturbation and its first conformal time derivative 
continuously across the 
inflation-radiation dominated transition value $\eta_1$. 
\bes \label{Mam1}
{\psi}_I(\bar{\bar{\eta}}=\eta_1) & = & {\psi}_{II}(\bar{\eta}=\eta_1) \\
\left. {\frac{d{\psi}_I}{d{\bar{\bar{\eta}}}}}\right|_{\bar{\bar{\eta}} 
\; = \; \eta_1} & = &  \label{Mam2}
\left. {\frac{d{\psi_{II}}}{d{\bar{\eta}}}}\right|_{\bar{\eta} \; = \; \eta_1}
\ees
The derivative of the reduced amplitude perturbation can be computed as:
\bes
\frac{d{\psi}_I}{d{\bar{\bar{\eta}}}}
& = & - \frac{1}{2}{\mathcal{N}}\; 
{\left(\eta_I-
\bar{\bar{\eta}}\right)}^{-\frac{1}{2}}\;
{{\mathcal{H}}_{\nu}}^{(1)}\left(k(\eta_I-\bar{\bar{\eta}})\right) \: + 
\nonumber \\
&  & - \: {\mathcal{N}} k {\left(\eta_I-\bar{\bar{\eta}}\right)}^{\frac{1}{2}}
\frac{d{{\mathcal{H}}_{\nu}}^{(1)}\left(k(\eta_I-\bar{\bar{\eta}})\right)}
{d\left(k(\eta_I-\bar{\bar{\eta}})\right)} \\
\frac{d{\psi_{II}}}{d{\bar{\eta}}} 
& = & i {\mathcal{M}}  
\sqrt{\frac{2}{\pi k}}
{\left((-ik) \alpha  e^{-i k(\bar{\eta} - \eta_1)} + (ik) \beta 
e^{i k(\bar{\eta}-\eta_1)}
\right)}
\ees
In the next, we will call:
\be \label{hpri}
{{\mathcal{H'}}_{\nu}}^{(\;i\;)} \; = \; 
\left. \frac{d{{\mathcal{H}}_{\nu}}^{(\;i\;)}
\left(k(\eta_I-\bar{\bar{\eta}})\right)}
{d(\left(k(\eta_I-\bar{\bar{\eta}})\right))}\right|_{{\bar{\bar{\eta}}}=\eta_1}
\ee

In particular, the argument of the Hankel functions involved in the 
matching can be expressed, 
following eq. (\ref{MAT1})
\be \label{X:A}
{\left.k({\eta_I}-{\bar{\bar{\eta}}})\right|}_{\bar{\bar{\eta}}
\;=\;{\eta}_1} \: = \: k\; {\eta}_1 \; \frac{q}{r} \: = \: X
\ee

With this property, equations (\ref{Mam1}) and (\ref{Mam2}) can be 
written as:
\bes
{\mathcal{N}}{\left(X \right)}^{\frac{1}{2}}
{{\mathcal{H}}_{\nu}}^{(1)}\left(X \right) = 
i {\mathcal{M}}  \sqrt{\frac{2}{\pi}} (\alpha  + \beta)  & &
\label{M:I} \\
\frac{1}{2 k}{\mathcal{N}}{\left(X \right)}^{-\frac{1}{2}}
{{\mathcal{H}}_{\nu}}^{(1)}\left(X \right)  + 
k  {\mathcal{N}}{\left(X \right)}^{\frac{1}{2}}
{{\mathcal{H'}}_{\nu}}^{(1)}\left(X \right)
 =  {\mathcal{M}}  \sqrt{{2 \pi}}  (\beta  -  \alpha) & &
\label{M:II}
\ees

By solving this system, we find for the Bogoliubov's 
coefficients:
\bes
\alpha & = & - \; \sqrt{\frac{\pi}{2}} \: \frac{\mathcal{N}}{2 \mathcal{M}}
{X}^{\frac{1}{2}}
\left[\left(\frac{1}{2X} \; + \; i \;\right){{\mathcal{H}}_{\nu}}^{(1)}
(\;X\;) \;+\;
{{\mathcal{H'}}_{\nu}}^{(1)}(\;X\;)\right]  \label{alfa2} \\
\beta & = & \sqrt{\frac{\pi}{2}} \; \frac{\mathcal{N}}
{2 \mathcal{M}}{X}^{\frac{1}{2}}
\left[\left(\frac{1}{2X} \; - \; i \;\right){{\mathcal{H}}_{\nu}}^{(1)}
(\;X\;) \;+\;
{{\mathcal{H'}}_{\nu}}^{(1)}(\;X\;) \right] \label{beta2}
\ees

From the properties of the Hankel functions (See \cite{tag}),
and their conjugated, for real $\nu$ and $X$ which is the
case here:
\bes
\overline{{{\mathcal{H}}_{\nu}}^{(1)}(\;X\;)} & = & {{\mathcal{H}}_{\nu}}^{(2)}
(\;X\;) \\
\overline{{{\mathcal{H'}}_{\nu}}^{(1)}(\;X\;)} & = & 
{{\mathcal{H'}}_{\nu}}^{(2)}
(\;X\;) 
\ees
The conjugate of Bogoliubov's coefficients 
$\bar{\alpha}$ and 
$\bar{\beta}$ can be written as:
\bes
\bar{\alpha}& = & - \; \sqrt{\frac{\pi}{2}} \; \frac{\bar{\mathcal{N}}}
{2 \bar{\mathcal{M}}}
{X}^{\frac{1}{2}}
\left[\left(\frac{1}{2X}  -  i \;\right){{\mathcal{H}}_{\nu}}^{(2)}
(\;X\;) \;+\;
{{\mathcal{H'}}_{\nu}}^{(2)}(\;X\;)\right]  \label{alfa3} \\
\bar{\beta} & = & \sqrt{\frac{\pi}{2}} \; \frac{\bar{\mathcal{N}}}
{2 \bar{\mathcal{M}}}
{X}^{\frac{1}{2}}
\left[\left(\frac{1}{2X}  +  i \;\right){{\mathcal{H}}_{\nu}}^{(2)}
(\;X\;) \;+\;
{{\mathcal{H'}}_{\nu}}^{(2)}(\;X\;) \right] \label{beta3}
\ees

With the properties of  Hankel functions (See \cite{tag}) and their 
wronskian (See \cite{taa}): 
\be \label{phwr}
\left\{{{\mathcal{H}}_{\nu}}^{(1)}(z),\; 
{{\mathcal{H}}_{\nu}}^{(2)}(z)\right\} 
\; = \; - i \; \frac{4}{\pi z}
\ee
and by rearranging coefficients,
we obtain the expressions for the square
modulii of Bogoliubov's 
coefficients.
\bes
{\mid\alpha\mid}^2  
& = & \frac{\pi}{8}\frac{{\mid{\mathcal{N}}\mid}^2}{{\mid{\mathcal{M}}\mid}^2}
\frac{{\left(\nu-\frac{1}{2}\right)}^2}{X}
\left[ \left(1+\frac{X^2}{{\left(\nu-\frac{1}{2}\right)}^2}\right)
{{\mathcal{H}}_{\nu}}^{(1)}(X){{\mathcal{H}}_{\nu}}^{(2)}(X) \; + \;
\right. \nonumber \\
& + & \frac{X^2}{{\left(\nu-\frac{1}{2}\right)}^2} \;
{{\mathcal{H}}_{\nu-1}}^{(1)}(X)
{{\mathcal{H}}_{\nu-1}}^{(2)}(X) \; + \; \nonumber \\
& - & \mbox{}\left. \frac{2X}{\left(\nu-\frac{1}{2}\right)}
{{\mathcal{H}}_{\nu}}^{(1)}(X){{\mathcal{H}}_{\nu-1}}^{(2)}(X) 
 - i \frac{4}{\pi \left(\nu-\frac{1}{2}\right)}  +  \frac{4\;X}
{\pi {\left(\nu-\frac{1}{2}\right)}^2} \right] \label{Alfa}
\ees
\bes
{\mid\beta\mid}^2 
& = & \frac{\pi}{8}\frac{{\mid{\mathcal{N}}\mid}^2}{{\mid{\mathcal{M}}\mid}^2}
\frac{{\left(\nu-\frac{1}{2}\right)}^2}{X}
\left[ \left(1+\frac{X^2}{{\left(\nu-\frac{1}{2}\right)}^2}\right)
{{\mathcal{H}}_{\nu}}^{(1)}(X){{\mathcal{H}}_{\nu}}^{(2)}(X) \;+\;
\right. \nonumber \\
& + & \frac{X^2}{{\left(\nu-\frac{1}{2}\right)}^2} \; 
{{\mathcal{H}}_{\nu-1}}^{(1)}(X)
{{\mathcal{H}}_{\nu-1}}^{(2)}(X) \; + \; \nonumber \\
& - & \mbox{}\left. \frac{2X}{\left(\nu-\frac{1}{2}\right)}
{{\mathcal{H}}_{\nu}}^{(1)}(X){{\mathcal{H}}_{\nu-1}}^{(2)}(X) 
 - i \frac{4}{\pi \left(\nu-\frac{1}{2}\right)}  -  \frac{4\;X}
{\pi {\left(\nu-\frac{1}{2}\right)}^2} \right] \label{Beta}
\ees

By requiring the Bogoliubov's 
coefficients satisfy the
usual normalization condition:
\be \label{bog}
{\mid\alpha\mid}^2-{\mid\beta\mid}^2 = 1
\ee
we have the constraint:
\be \label{nm}
\frac{{\mid{\mathcal{N}}\mid}^2}{{\mid{\mathcal{M}}\mid}^2} \: = \: 1
\ee

This condition holds for all $k$, thus it ensure us the absence of 
isolated dependence of $k$ 
through the normalization constants.
The difference among them is only a phase.
The same result can be reached by computing explicitly the constants 
${\mathcal{N}}$ and
${\mathcal{M}}$ in the framework of a specific normalization scheme. 

\subsubsection{The Power Spectrum in Physical Units}
 
\margen Now we are able to compute the power spectrum of the stochastic 
gravitational wave background 
with full physical meaning. 
In terms of the proper frequency $\omega$ at detection time 
eq.(\ref{deom}), we read the
variable $X$ eq.(\ref{X:A}) as:
\be
X \; = \; \omega \; \left(\frac{q}{r}\right)\frac{a(\eta_0) \eta_1}{c}
\ee

Notice that the dependence with the global scale factor 
parameter $\bar{A_{II}}$
(the only quantity unknown in these expressions),
is cancelled in $X$ by the product of  $\eta_1 a(\eta_0)$. 
With the help of eqs.(\ref{et1ob}), (\ref{Ahoy}), (\ref{SET1}) 
and some rearrangements
we have:
\be \label{ese}
X \equiv \omega \; S \ \ \ \ , \ \ \ S \; = \;  \frac{Q}{(Q+1)M}
{\left(\frac{{\mathcal{T}}_0}{{\mathcal{T}}_m}\right)}^M
{\left(\frac{{\mathcal{T}}_m}{{\mathcal{T}}_r}\right)}^R\;{\mathcal{T}}_r
\ee

That is, we find the argument $X$ 
related to the physical proper frequency in a way 
{\it totally determined} by the 
cosmological model studied, and linked to 
observational times.
The ratii ${\left(\frac{{\mathcal{T}}_0}{{\mathcal{T}}_m}\right)}^M$ and    
${\left(\frac{{\mathcal{T}}_m}{{\mathcal{T}}_r}\right)}^R$  are exactly 
the scale
factor expansion reached during the radiation dominated and matter 
dominated epochs, respectively, introduced as part of our linking
to observational Universe information on the minimal cosmological model. 
(See section III). 
Both values are strongly constrained 
by the observational evolution data. Meanwhile,  the value 
${{\mathcal{T}}_r}$
(beginning of radiation dominated stage) gives us the order of magnitude 
of the exit
of inflationary stage, related to the order of magnitude of GUT scales. 

Obviously, the quantity $S$ has dimensions $[S] = {\mathrm{time}}$. 
Thus, as $[\omega] = {\mathrm{rad}}\;{\mathrm{(time)}}^{-1}$, 
$X$ is dimensionless.  We can give here the value of $S$ for our cosmological 
scale factor. From String Theory, we have obtained the scale factors with 
parameters of temporal
dependence $Q=\frac{2}{d+1}$, $R=\frac{2}{d+1}$ and $M=\frac{2}{d}$ 
for inflationary,
radiation dominated and matter dominated stages, respectively 
(See eqs.(\ref{sdrin}), 
(\ref{sdrad}), (\ref{rgmat}) and table(\ref{t.cos})). By linking with the 
cosmological observational description,
we have considered the standard values for  ${{\mathcal{T}}_r}$, 
${{\mathcal{T}}_m}$
and ${{\mathcal{T}}_0}$ given in previous section. This leads to the expression
\bef
X_{String} \; = \; \omega \;\frac{d}{(3+d)}\;{\left(\frac{{\mathcal{T}}_0}
{{\mathcal{T}}_m}\right)}^{\frac{2}{d}}
\;{\left(\frac{{\mathcal{T}}_m}{{\mathcal{T}}_r}\right)}^{\frac{2}{d+1}}
\;{\mathcal{T}}_r \ \ \ \ .
\eef
For three spatial 
dimensions, and the time
constants given in seconds, we have:
\be \label{x3}
X_{String} \; = \; \omega \; \frac{1}{2} 10^{-\frac{20}{3}} {\mathrm{seg }}
\ee  
If we define a characteristic frequency $\omega_x$ which makes unity 
the argument $X$ 
($\omega_x \; S \; = \; 1 $), the value of it would be
\be \label{omx}
\omega_x \; \sim \;  9.28 \; 10^6 \; \frac{\mathrm{rad}}{\mathrm{s}} \: 
\sim  \: 1.48 \; 
{\mathrm{Mhz}}
\ee

Now, we compute the full and exact expression of the power spectrum for the 
stochastic background of
gravitational waves produced in the inflation-radiation dominated 
transition. Following the usual definition, the power spectrum of 
gravitational waves computed in three spatial 
dimensions is: 
\be
P(\omega) d\omega = \frac{\hbar}{{\pi}^2 c^3}{\omega}^3 d\omega 
{\mid\beta\mid}^2
\ee
From eq.(\ref{Beta}) and eq.(\ref{ese}), we have:
\bes \label{P:All}
P(\omega) d\omega & = & \frac{\hbar}{8 \pi c^3}{\omega}^2  
\frac{{\left(\nu-\frac{1}{2}\right)}^2}{S} d\omega
\left[ \left(1+\frac{(\omega S)^2}{{\left(\nu-\frac{1}{2}\right)}^2}\right)
{{\mathcal{H}}_{\nu}}^{(1)}(\omega S){{\mathcal{H}}_{\nu}}^{(2)}(\omega S) 
\;+\;
\right. \nonumber \\
& + & \frac{(\omega S)^2}{{\left(\nu-\frac{1}{2}\right)}^2} \; 
{{\mathcal{H}}_{\nu-1}}^{(1)}(\omega S)
{{\mathcal{H}}_{\nu-1}}^{(2)}(\omega S) \; + \; \nonumber \\
& - & \left. \frac{2 \; \omega S}{\left(\nu-\frac{1}{2}\right)}
{{\mathcal{H}}_{\nu}}^{(1)}(\omega S){{\mathcal{H}}_{\nu-1}}^{(2)}(\omega S) 
 -  \frac{4\;\omega S}
{\pi {\left(\nu-\frac{1}{2}\right)}^2}- i 
\frac{4}{\pi \left(\nu-\frac{1}{2}\right)} \right] 
\ees
where $S$ is given by eq.(\ref{ese}).

The expression in square brackets is dimensionless. All physical units 
are found in the term
multiplying it and proportional to $\omega^2 d\omega$. For our scale 
factor in three spatial 
dimensions, the coefficient in front will take the value: 
\be \label{cfp}
\frac{\hbar}{8 \pi c^3}\; \frac{{\left(\nu-\frac{1}{2}\right)}^2}{S} \: 
\sim \: 
1.603 \; 10^{-54} \; \frac{{\mathrm{erg}}\;{\mathrm{seg}}^3}{{\mathrm{cm}}^3}
\ee
thus the power spectrum $P(\omega) d \omega$ has 
dimensions of energy density.

Notice the dependence of $P(\omega)$ on the scale factor
parameters through the parameter $\nu$, eq.(\ref{nu:A}), 
characterizing the expression of the power spectrum 
and its $\omega$ dependence. This is the way the
cosmological theory considered is
imprinted on the gravitational wave spectrum.
The parameter X eq.(\ref{ese}), is uniquely fixed by the  
scale factor and according with observational values. We will
see in the next sections as the different dilaton roles and the different
theories, although involving the same scale factor,
led to different power spectra shapes. 

It is also possible to express the power spectrum
in terms of the Bessel functions ${{\mathcal{J}}_{\nu}}(z)$
and the Neumann functions 
${{\mathcal{Y}}_{\nu}}(z)$  \cite{tag}:
\bes
{{\mathcal{H}}_{\nu}}^{(1)}(z) & = & {{\mathcal{J}}_{\nu}}(z) + i 
{{\mathcal{Y}}_{\nu}}(z) \\
{{\mathcal{H}}_{\nu}}^{(2)}(z) & = & {{\mathcal{J}}_{\nu}}(z) - i 
{{\mathcal{Y}}_{\nu}}(z)
\ees
with the Wronskian\cite{taa}
\be \label{jnwr}
\left\{{{\mathcal{J}}_{\nu}}(z),\; 
{{\mathcal{Y}}_{\nu}}(z)\right\} \;  = \; \frac{2}{\pi z} \ \ .
\ee

\pagebreak[4]

The power spectrum is expressed as:
\bes{\label{Py:All}}
P(\omega) d\omega & = & \frac{\hbar}{8 \pi c^3}{\omega}^2  
\frac{{\left(\nu-\frac{1}{2}\right)}^2}{S} d\omega \nonumber \\
& & \left[\left(1+\frac{(\omega S)^2}{{\left(\nu-\frac{1}{2}\right)}^2}\right)
{\left(\frac{\mbox{}}{\mbox{}}{\mathcal{J}}_{\nu}(\omega S)
{\mathcal{J}}_{\nu}(\omega S) + 
{\mathcal{Y}}_{\nu}(\omega S){\mathcal{Y}}_{\nu}(\omega S)\right)} 
\right. + \nonumber \\
& + & \frac{(\omega S)^2}{{\left(\nu-\frac{1}{2}\right)}^2} \; 
{\left(\frac{\mbox{}}{\mbox{}}{\mathcal{J}}_{{\nu}-1}(\omega S)
{\mathcal{J}}_{{\nu}-1}(\omega S) 
+ {\mathcal{Y}}_{{\nu}-1}(\omega S){\mathcal{Y}}_{{\nu}-1}(\omega S)\right)} 
+ \nonumber \\ 
& - & \frac{2 \; \omega S}{\left(\nu-\frac{1}{2}\right)}
{\left(\frac{\mbox{}}{\mbox{}}{\mathcal{J}}_{\nu}(\omega S)
{\mathcal{J}}_{{\nu}-1}(\omega S) 
+ {\mathcal{Y}}_{\nu}(\omega S){\mathcal{Y}}_{{\nu}-1}
(\omega S)\right)} + \nonumber \\
& - & \left. \frac{4\;\omega S}
{\pi {\left(\nu-\frac{1}{2}\right)}^2} \right]
\ees

Another expression for the power spectrum is in terms
of only Bessel functions ${\mathcal{J}}_{\nu}$, ${\mathcal{J}}_{-\nu}$.
In order to make this, we use the relations (\cite{tag}, \cite{taa}):
\besf
{{\mathcal{H}}_{\nu}}^{(1)}(z) & = & \frac{i}{\sin(\pi \nu)} 
{\left({\mathcal{J}}_{\nu}(z) e^{-i \pi \nu} - 
{{\mathcal{J}}_{-\nu}}(z)\right)} \\
{{\mathcal{H}}_{\nu}}^{(2)}(z) & = & \frac{i}{\sin(\pi \nu)} 
{\left({\mathcal{J}}_{-\nu}(z) - 
e^{i \pi \nu}{{\mathcal{J}}_{\nu}}(z)\right)} \\
\left\{{{\mathcal{J}}_{\nu}}(z),\; 
{{\mathcal{J}}_{-\nu}}(z)\right\} & = &
 - \frac{2 \; \sin(\pi \nu)}{\pi z}
\eesf
We find the next expression dealing only Bessel functions:
\bes \label{Pjj:All}
\lefteqn{P(\omega) d\omega =  \frac{\hbar}{8 \pi c^3}{\omega}^2  
\frac{{\left(\nu-\frac{1}{2}\right)}^2}{S \; \sin^2(\pi\nu)} 
d\omega} & &  \nonumber \\
& & \left[\left(1+\frac{(\omega S)^2}{{\left(\nu-\frac{1}{2}\right)}^2}\right)
\left(\frac{\mbox{}}{\mbox{}}{\mathcal{J}}_{\nu}(\omega S)
{\mathcal{J}}_{\nu}(\omega S) + 
{\mathcal{J}}_{-\nu}(\omega S){\mathcal{J}}_{-\nu}(\omega S) + 
\right. \right. \nonumber \\
& & \left. - \; 2 \cos({\pi \nu}) {\mathcal{J}}_{\nu}(\omega S)
{\mathcal{J}}_{-\nu}(\omega S) \frac{\mbox{}}{\mbox{}}\right) +  \nonumber \\
& + & \frac{(\omega S)^2}{{\left(\nu-\frac{1}{2}\right)}^2}
\left(\frac{\mbox{}}{\mbox{}} {\mathcal{J}}_{\nu-1}(\omega S)
{\mathcal{J}}_{\nu-1}(\omega S) + 
{\mathcal{J}}_{1-\nu}(\omega S){\mathcal{J}}_{1-\nu}(\omega S) +
\right. \nonumber \\ 
& & \left. + \; 2 \cos({\pi \nu}) {\mathcal{J}}_{\nu-1}(\omega S)
{\mathcal{J}}_{1-\nu}(\omega S)\frac{\mbox{}}{\mbox{}} 
\right) + \nonumber \\
& - & \frac{2 \; \omega S}{\left(\nu-\frac{1}{2}\right)} 
{\left(\frac{\mbox{}}{\mbox{}}{\mathcal{J}}_{\nu}(\omega S)
{\mathcal{J}}_{\nu-1}(\omega S) - 
{\mathcal{J}}_{-\nu}(\omega S){\mathcal{J}}_{1-{\nu}}(\omega S)
\frac{\mbox{}}{\mbox{}} \right)} 
+ \nonumber \\
& - &  \frac{2 \; \omega S \; \cos{(\pi \nu)}}{\left(\nu-\frac{1}{2}\right)}
{\left( \frac{\mbox{}}{\mbox{}} {{\mathcal{J}}_{\nu}}(\omega S)
{\mathcal{J}}_{1-{\nu}}(\omega S) - {\mathcal{J}}_{-\nu}(\omega S) 
{\mathcal{J}}_{\nu-1}(\omega S)\right)} + \nonumber \\
& - & \left. \frac{4\;\omega S \; {\sin^2{(\pi\;\nu)}}}
{\pi {\left(\nu-\frac{1}{2}\right)}^2} \right]
\ees

Eqs.(\ref{Pjj:All}) and (\ref{Py:All}) manifest explicitely the 
real character of the mathematical expressions involved in the power 
spectrum. In fact, the imaginary terms belonging to 
Hankel functions cancel exactly among them and with the free 
imaginary term.
This is more easily seen looking in the last expression, where
the power spectrum is fully expressed in Bessel functions 
(real functions when the argument is real) (See \cite{taj})
From the numerical point of view, 
expressions (\ref{Py:All}) or
(\ref{Pjj:All}) are more convenient, since standard algorithms are
limited in compute Hankel functions to very high precision at 
large arguments and it leds to wrong results in computing the
above expressions.
 Expressions (\ref{Py:All}) and (\ref{Pjj:All}) 
allow extend 1-2 orders of magnitude more the numerical
computations of the exact spectrum. In any case, in the next
chapter we will overcome this limitation with analytical 
expressions for the regime of  small and large arguments.

Finally, we compute the fraction of critical energy density in 
gravitational waves by octave. Following the definition 
eq.(\ref{ome}) and using the expression (\ref{P:All}), we have:
\bes \label{om:All}
\Omega_{GW} & = & \frac{\hbar \; G}{3 {H_0}^2 c^5}\;
\frac{{\left(\nu-\frac{1}{2}\right)}^2}{ S} \; \omega^3
\left[ \left(1+\frac{(\omega S)^2}{{\left(\nu-\frac{1}{2}\right)}^2}\right)
{{\mathcal{H}}_{\nu}}^{(1)}(\omega S){{\mathcal{H}}_{\nu}}^{(2)}(\omega S) 
\;+\; \right. \nonumber \\
& + & \frac{(\omega S)^2}{{\left(\nu-\frac{1}{2}\right)}^2} \; 
{{\mathcal{H}}_{\nu-1}}^{(1)}(\omega S)
{{\mathcal{H}}_{\nu-1}}^{(2)}(\omega S) \; + \; \nonumber \\
& - & \left. \frac{2 \; \omega S}{\left(\nu-\frac{1}{2}\right)}
{{\mathcal{H}}_{\nu}}^{(1)}(\omega S){{\mathcal{H}}_{\nu-1}}^{(2)}(\omega S) 
 -  \frac{4\;\omega S}
{\pi {\left(\nu-\frac{1}{2}\right)}^2}- i 
\frac{4}{\pi \left(\nu-\frac{1}{2}\right)} \right] 
\ees

 Again, we will compute the value of the coefficient multiplying $\omega^3$
and the expression in square brackets:
\be \label{cfo}
\frac{\hbar \; G}{3 {H_0}^2 c^5} \; \frac{{\left(\nu-\frac{1}{2}\right)}^2}
{S} \: \sim \: 2.488 \; 10^{-46} \; {\mathrm{seg}}^{-3}
\ee

\subsection{Partial Dilaton Case}

\margen In this case, we will consider the role of the dilaton field in the 
equations for the reduced
amplitude perturbation $\psi$. We will see that 
this procedure still means a
partial consideration of the dilaton role on the gravitational wave 
spectra. By now, we will solve eqs.(\ref{eq:SD}) and (\ref{V:SD}) 
with the reduced
amplitude perturbation given by (\ref{psi:SD}).

Now, in the inflationary stage, we take into account not only the 
scale factor but
also the dilaton field. Following eqs.(\ref{sdrin})
and (\ref{EDescr})
we have for the dilaton in this stage:
\bes \label{dil}
\phi(\bar{\bar{\eta}}) & = & \phi_I + 2d \ln a(\bar{\bar{\eta}})  
\ees

Thus, the potential eq.(\ref{V:SD}) takes now the form:
\be \label{V:pdi}
V(\eta) \; = \; \left({\left(\frac{d+1}{2}q-\frac{1}{2}\right)}^2
-\frac{1}{4}\right){(\eta_I-\bar{\bar{\eta}})}^{-2}
\ee

Then, we have again an equation whose solution is expressed
in terms of Bessel functions,
but with the parameter $\nu$ having the value:
\be \label{nu:SD}
\nu \; = \; \pm \; {\left(\frac{d+1}{2}q - \frac{1}{2}\right)}
\ee
Or, expressed in terms of the spatial number dimensions,
\be
\nu \: = \: \frac{d-1}{2(d+3)}
\ee
that is, $\nu \; = \; \frac{1}{6}$ for three spatial dimensions,
and fixing the positive sign in the eq.(\ref{nu:SD}).

Notice the difference with the parameter $\nu$ eq.(\ref{nu:ND})
obtained in the No Dilaton Case. 
The difference among both cases is:
\bef
\left.\nu \right|_{{\mathrm{part. dil.}}} \; = \; 
\left.\nu\right|_{{\mathrm{no\;dil.}}}\;+\, (1- q)
\eef

It leds to a totally similar expression to
eq.(\ref{vacinf}) for the 
solution of the reduced amplitude perturbation:
\be 
{\psi}_I \: = \: {\mathcal{N}} {\left(\eta_I-
\bar{\bar{\eta}}\right)}^{\frac{1}{2}}
{{\mathcal{H}}_{\nu}}^{(1)}\left(k(\eta_I-\bar{\bar{\eta}})\right)
\ee
where $\nu$ is given now by eq.(\ref{nu:SD})
and ${\mathcal{N}}$ is a normalization constant.

For the radiation dominated stage, the scale factor is given by
eq.(\ref{EDescr}) and the dilaton field, according to eqs.(\ref{sdrad}),
remains "frozen" in its constant value at the exit inflation time:
\be \label{dilrr}
\phi_{II} \; = \; \left.\phi(\bar{\bar{\eta}})\right|_{\bar{\bar{\eta}}
\;=\;\eta_1} 
\ee

Thus, we recover exactly the same expressions for the potential 
eq.(\ref{V:ndr}), the general solution eq.(\ref{radgen}), the
parameter $\mu=\pm\frac{1}{2}$ and consequently, the 
same expression in terms of Bogoliubov's coefficients
$\alpha$ and $\beta$:
\be \label{psirad}
{\psi}_{II}  =  i {\mathcal{M}}  \sqrt{\frac{2}{\pi k}}
{\left(\alpha e^{-i k(\bar{\eta} - \eta_1)} + \beta e^{i k(\bar{\eta}-\eta_1)}
\right)}
\ee
with ${\mathcal{M}}$ a normalization constant.

The Bogoliubov's coefficients computed now 
have exactly the same formal expressions yet obtained in the
last section, 
eqs.(\ref{Alfa}) and (\ref{Beta}), since the reduced amplitude
perturbations involved in the matching at the transition
at $\eta_1$ are formally the same functions
(\ref{vacinf}) and (\ref{vacrad}) in both cases. Only the
parameter $\nu$ is different, now given by eq.(\ref{nu:SD}).

Similarly, all the relations in order to compute the power spectrum 
eq.(\ref{P:All}), the fraction of critical energy density 
$\Omega_{GW}$ eq.(\ref{om:All}) and the expression of the argument 
$X$ eq.(\ref{ese}) as function of the proper frequency $\omega$, hold in this
case.  Notice that the argument $X$ does not depend on the dilaton role; 
this quantity comes fixed from the evolution 
of the scale factor. Thus, we have again $P(\omega)
d \omega$ and $\Omega_{GW}$ as given by eqs.(\ref{P:All}) and
(\ref{om:All}) but with the parameter $\nu$ given in this case
by eq.(\ref{nu:SD}).

In three spatial dimensions, 
we have the same values for the
parameter $S$ eq.(\ref{x3}) and for the characteristic
frequency $\omega_x$ eq.(\ref{omx}). 
The numerical values of coefficients multiplying the square
brackets expressions in (\ref{P:All}) and (\ref{om:All}) would
be, in principle, differents since they depend of the parameter
$\nu$. But in our case, in three spatial dimensions, the
value of $(\nu-\frac{1}{2})^2$ is the same for both cases,
giving the same numerical value for these coefficients too.

It must be observed that the normalization constants ${\mathcal{N}}$
and ${\mathcal{M}}$ are constrainted by the normalization condition of 
Bogoliubov's coefficients to satisfy the same relation found in the case
before: ${\mathcal{N}}$ and ${\mathcal{M}}$ differ only in a phase. 
(See eqs.(\ref{nm})). But the case
of particular and explicit normalization applicable to the no dilaton case
is not generalizable to the dilaton case. Now, both constants will
remain as undetermined parameters in the amplitude perturbations.

\subsection{The Full Dilaton Role}

\margen In this section, we will take into account the full dilaton role.
In the last section, we have considered it through the reduced
amplitude perturbation $\psi$, solution of eq.(\ref {eq:SD}).
But we have left unconsidered the dilaton role in the total
amplitude perturbation $h_{\mu \nu}$. In order to include this, we will
match in this case the metric amplitude perturbation $h_{\mu\nu}$
instead of only the reduced amplitude perturbation $\psi$.

In the framework of General Relativity, as the treatment of
Grishchuk \cite{gks} and Allen \cite{A}, both procedures
are equivalents, since the metric perturbation and
the reduced amplitude differ only through a power of the 
scale factor. In those cases, and with a description with suitable 
properties of continuity for the scale factor and its first 
conformal time derivative like those we are dealing with, the 
final result for the gravitational wave spectra would be
the same, since there are no lack of information by
matching only a part of the metric perturbation, the reduced
amplitude. The remaining part would give us redundant information.

But in the case of theories with dilaton fields, Brans-Dicke frames 
and in particular, string cosmology, 
the remaining part in the total amplitude perturbation contains 
additional information with respect to the 
reduced one. It contains an added dependence on
the dilaton field, as can be seen from eq.(\ref{hm:SD}). The behaviour of
this part through the transition is not the same as the scale factor, since
there is no continuity in the first derivative of the dilaton
field. Thus, the power spectra computed by matching only the
reduced amplitude perturbations is different from that obtained by matching
the full metric perturbation. 

In summary, the physical quantity
to be studied is the full  metric perturbation amplitude. Meanwhile,
the reduced amplitude perturbation 
can be considered as a sometimes convenient 
mathematical reduction of the problem.

In the inflationary stage, we have
the same equations for the reduced amplitude
perturbation (\ref{eq:SD}) and (\ref{V:SD}) that in the 
precedent section. Therefore, we will have the same expression
for the reduced amplitude perturbation:
\be \label{vacFD}
{\psi}_I \: = \: {\mathcal{N}} {\left(\eta_I-
\bar{\bar{\eta}}\right)}^{\frac{1}{2}}
{{\mathcal{H}}_{\nu}}^{(1)}\left(k(\eta_I-\bar{\bar{\eta}})\right)
\ee
with
\be \label{nu:FD}
\nu \; = \; \pm \; {\left(\frac{d+1}{2}q - \frac{1}{2}\right)}
\ee

The reduced $\psi$ and the total tensorial 
metric perturbation are related by eq.(\ref{psi:SD}). Instead of 
$\psi$, we will match the metric
perturbation amplitude $\Phi$:
\be \label{Phi}
\Phi \: = \: \psi a^{-\frac{d-1}{2}} e^{\frac{\phi}{2}}
\ee
such that the metric perturbation is written as:
\be \label{hmn}
h_{\mu \nu} \; = \; e_{\mu \nu}({\bf k})\;\Phi(\eta,{\bf k})\; e^{i\bf{k.x}}
\ee
From eqs.(\ref{sdrin}), (\ref{EDescr})
and (\ref{dil}), we have:
\be
\Phi_I \; =  \; {\mathcal{N}}{\alpha_I}^{\frac{d+1}{2}}
e^{\frac{\phi_I}{2}}\;{\left(\eta_I-\bar{\bar{\eta}}
\right)}^{-q\left(\frac{d+1}{2}\right)+\frac{1}{2}}
{\mathcal{H}_{\nu}}^{(1)}\left(k(\eta_I-\bar{\bar{\eta}})\right)
\ee
In contrast to the precedent cases, we will take in the
following the negative sign in the parameter $\nu$ (\ref{nu:FD}):
\be \label{nu:FD2}
\nu \; =  {\left(\frac{1}{2}\;-\; q \frac{d+1}{2}\right)}
\ee
because with this definition, we can write the amplitude 
perturbation as:
\be \label{phin}
\Phi_I \; =  \; {\mathcal{N}}{\alpha_I}^{\frac{d+1}{2}}
e^{\frac{\phi_I}{2}}\;{\left(\eta_I-\bar{\bar{\eta}}\right)}^{\nu} 
{\mathcal{H}_{\nu}}^{(1)}\left(k(\eta_I-\bar{\bar{\eta}})\right)
\ee
where ${\mathcal{N}}$ is a normalization constant.Notice the 
parameter $\nu$ in the conformal time dependence.

On the other hand, in the radiation dominated stage,
the reduced amplitude perturbation is
obtained in a totally equivalent way to the precedent section.
We recover for this reduced amplitude
eq.(\ref{psirad}):
\be 
{\psi}_{II}  =  i {\mathcal{M}}  \sqrt{\frac{2}{\pi k}}
{\left(\alpha e^{-i k(\bar{\eta} - \eta_1)} + \beta 
e^{i k(\bar{\eta}-\eta_1)}
\right)}
\ee
since we have again the parameter $\mu=\pm\frac{1}{2}$.

The tensorial metric perturbation $h_{\mu \nu}$ is given by 
eq.(\ref{hmn}) with the amplitude $\Phi$ given by:
\bes \label{phirad}
\Phi_{II} & = &  i {\mathcal{M}}  \sqrt{\frac{2}{\pi k}} 
e^{\frac{\phi_{II}}{2}}
{\alpha_{II}}^{-\frac{d-1}{2}} {\bar{\eta}}^{-r\left(\frac{d-1}{2}\right)}
{\left(\alpha e^{-i k(\bar{\eta} - \eta_1)} + \beta e^{i k(\bar{\eta}-\eta_1)}
\right)}
\ees
where $r=\frac{2}{d-1}$.
In shorter form, introducing the notation:
\bes
{\mathcal{F}} & = & {\mathcal{N}}{\alpha_I}^{\frac{d+1}{2}}
e^{\frac{\phi_I}{2}}  \label{F:FD}\\
{\mathcal{G}} & = & i \sqrt{\frac{2}{\pi}}\;{\mathcal{M}}\;
{\alpha_{II}}^{-\frac{d-1}{2}} e^{\frac{\phi_{II}}{2}} \label{G:FD} \  \  ,
\ees
we have:
\bes \label{phies}
\Phi_I & =  & {\mathcal{F}}{\left(\eta_I-\bar{\bar{\eta}}\right)}^{\nu} 
{\mathcal{H}_{\nu}}^{(1)}\left(k(\eta_I-\bar{\bar{\eta}})\right) \\
\Phi_{II} & = & {\mathcal{G}} k^{-\frac{1}{2}} 
{\bar{\eta}}^{-1}
{\left(\alpha e^{-i k(\bar{\eta} - \eta_1)} + \beta e^{i k(\bar{\eta}-\eta_1)}
\right)}
\ees
where $\nu$ is given by eq.(\ref{nu:FD}).

As indicated above, in this case we will compute the Bogoliubov's 
Coefficients by matching the total amplitude of  tensorial metric
perturbation $\Phi$. Thus, we have to solve the equations:
\bes \label{Maph1}
{\Phi}_I(\bar{\bar{\eta}}=\eta_1) & = & {\Phi}_{II}(\bar{\eta}=\eta_1) \\
\left. {\frac{d{\Phi}_I}{d{\bar{\bar{\eta}}}}}\right|_{\bar{\bar{\eta}} 
\; = \; \eta_1} & = & \label{Maph2} 
\left. {\frac{d{\Phi_{II}}}{d{\bar{\eta}}}}\right|_{\bar{\eta} \; = \; \eta_1} 
\ees
which yield:
\bes
{\mathcal{F}}{\left({\frac{q}{r}}\eta_1\right)}^{\nu}
{{\mathcal{H}}_{\nu}}^{(1)}\left(k\frac{q}{r} \eta_1\right) & = &  
{\mathcal{G}} k^{-\frac{1}{2}} {\eta_1}^{-1} (\alpha  + \beta) 
\label{Mph:I} \\
{\mathcal{F}}\nu {\left(\frac{q}{r} \eta_1\right)}^{\nu-1}
{{\mathcal{H}}_{\nu}}^{(1)}\left(k \frac{q}{r} \eta_1\right) & + &
k  {\mathcal{F}}{\left(\frac{q}{r} \eta_1\right)}^{\nu}
{{\mathcal{H'}}_{\nu}}^{(1)}\left(k \frac{q}{r} \eta_1\right) \ \  =
\label{Mph:II} \\
& =  & {\mathcal{G}} k^{-\frac{1}{2}} {\eta_1}^{-2}  
(\beta (1-ik\eta_1) + \alpha (1+ik\eta_1))  \nonumber
\ees

By solving this system, we have: 
\bes
\alpha & =  & - i  \frac{\mathcal{F}}{2{\mathcal{G}}} {\eta_1}^{\nu+1}
{\left(\frac{q}{r}\right)}^{\nu} k^{\frac{1}{2}}
\left[\left(\frac{\nu}{X}- \frac{\frac{q}{r}}{X} + i \right) 
{{\mathcal{H}}_{\nu}}^{(1)}{\left(X \right)} +
{{\mathcal{H'}}_{\nu}}^{(1)}{\left(X \right)}\right] \label{palf2} \\
\beta & = & i  \frac{\mathcal{F}}{2{\mathcal{G}}} {\eta_1}^{\nu+1}
{\left(\frac{q}{r}\right)}^{\nu} k^{\frac{1}{2}}
\left[\left(\frac{\nu}{X}- \frac{\frac{q}{r}}{X} - i \right) 
{{\mathcal{H}}_{\nu}}^{(1)}{\left(X\right)} +
{{\mathcal{H'}}_{\nu}}^{(1)}{\left(X \right)}\right] \label{pbet2}
\ees
where $X$ is given by eq.(\ref{X:A}).

Finally, after computation and using the properties of 
Hankel functions and their wronskian (See \cite{taa} and \cite{tag}), 
we have the
following expressions for the square moduli:
\bes
{\mid\alpha\mid}^2  
& = & \frac{{\mid{\mathcal{F}}\mid}^2}
{{\mid{\mathcal{G}}\mid}^2}\;{\eta_1}^{2 \nu+2}
{\left(\frac{q}{r}\right)}^{2\nu} \; \frac{k}{4} \;
\left[ \left(1+\frac{{\left(\frac{q}{r}\right)}^2}{X^2}\right)
{{\mathcal{H}}_{\nu}}^{(1)}(X){{\mathcal{H}}_{\nu}}^{(2)}(X) \;+\;
\right. \nonumber \\
& + & {{\mathcal{H}}_{\nu-1}}^{(1)}(X){{\mathcal{H}}_{\nu-1}}^{(2)}(X) 
\nonumber \\
& - & \mbox{}\left. 2 \frac{\frac{q}{r}}{X}
{{\mathcal{H}}_{\nu}}^{(1)}(X){{\mathcal{H}}_{\nu-1}}^{(2)}(X)  
 - i \frac{4}{\pi } \frac{\frac{q}{r}}{X^2} + 
\frac{4}{\pi  X} \right] \label{PAlfa}
\ees
\bes
{\mid\beta\mid}^2  
& = & \frac{{\mid{\mathcal{F}}\mid}^2}
{{\mid{\mathcal{G}}\mid}^2}\;{\eta_1}^{2 \nu+2}
{\left(\frac{q}{r}\right)}^{2\nu} \; \frac{k}{4} \;
\left[ \left(1+\frac{{\left(\frac{q}{r}\right)}^2}{X^2}\right)
{{\mathcal{H}}_{\nu}}^{(1)}(X){{\mathcal{H}}_{\nu}}^{(2)}(X) \;+\;
\right. \nonumber \\
& + & {{\mathcal{H}}_{\nu-1}}^{(1)}(X){{\mathcal{H}}_{\nu-1}}^{(2)}(X) 
\nonumber \\
& - & \mbox{}\left. 2 \frac{\frac{q}{r}}{X}
{{\mathcal{H}}_{\nu}}^{(1)}(X){{\mathcal{H}}_{\nu-1}}^{(2)}(X)  
 - i \frac{4}{\pi } \frac{\frac{q}{r}}{X^2} - 
\frac{4}{\pi  X} \right] \label{PBeta}
\ees
 
From expressions (\ref{PAlfa}) and (\ref{PBeta}) is easy
to see:
\be
{\mid\alpha\mid}^2-{\mid\beta\mid}^2 \; = \; 
\frac{{\mid{\mathcal{F}}\mid}^2}{{\mid{\mathcal{G}}\mid}^2}
\;{\eta_1}^{2\nu+2} \; {\left(\frac{q}{r}\right)}^{2 \nu}\; \frac{k}{4} \; 
\left[\frac{8}{\pi \;k \;\frac{q}{r} \; \eta_1}\right] 
\ee
which by applying the normalization condition of Bogoliubov's
Coefficients eq.(\ref{bog}), gives:
\be \label{fg1}
\frac{{\mid{\mathcal{F}}\mid}^2}{{\mid{\mathcal{G}}\mid}^2}
{\eta_1}^{2\nu+1} {\left(\frac{q}{r}\right)}^{2 \nu-1} \; 
\frac{2}{\pi} \; = \; 1
\ee

Now, we can translate this condition over the usual normalization constants
${\mathcal{N}}$ and ${\mathcal{M}}$. From the definitions
(\ref{F:FD}) and (\ref{G:FD}):
\bef
\frac{{\mid{\mathcal{F}}\mid}^2}{{\mid{\mathcal{G}}\mid}^2} \; = \;  
\frac{{\mid{\mathcal{N}}\mid}^2}{{\mid{\mathcal{M}}\mid}^2} \;
\frac{\pi}{2} \; \frac{e^{\phi_I}}{e^{{\phi}_{II}}} \;
\frac{{\alpha_I}^{d+1}}{{{\alpha}_{II}}^{1-d}}
\eef
From the matching properties of the scale factor in
conformal time eq.(\ref{MAT1}), we have:
\be \label{sfrap}
\frac{\alpha_I}{\alpha_{II}} \: = \: {\left(\frac{q}{r}\right)}^q\; 
{\eta_1}^{r+q}
\ee
And from eqs.(\ref{dili}) and (\ref{dilr}):
\be \label{drap}
\frac{e^{\phi_I}}{e^{{\phi}_{II}}} \; = \; {\alpha_{II}}^{-2d} \; 
{\eta_1}^{-2dr}
\ee
With eqs.(\ref{sfrap}) and (\ref{drap}) and the $\nu$ value 
eq.(\ref{nu:FD2}) we obtain:
\be \label{fg2}
\frac{{\mid{\mathcal{F}}\mid}^2}{{\mid{\mathcal{G}}\mid}^2} \; = \;  
\frac{{\mid{\mathcal{N}}\mid}^2}{{\mid{\mathcal{M}}\mid}^2} \;
\frac{\pi}{2} \; {\left(\frac{q}{r}\right)}^{1-2\nu}\;{\eta_1}^{1-2\nu+r(1-d)}
\ee
Since $r=\frac{2}{d-1}$, eqs.(\ref{fg1}) and (\ref{fg2}) yield:
\be \label{nmfd}
\frac{{\mid{\mathcal{N}}\mid}^2}{{\mid{\mathcal{M}}\mid}^2} \; = \;
{\eta_1}^{-2-r(1-d)} \: = \: 1
\ee

That is, the normalization factors
${\mathcal{N}}$ and ${\mathcal{M}}$ differ only in a phase.
We can eliminate the normalization constants in the
expressions for the Bogoliubov's Coefficients. From eq.(\ref{fg1}), 
the coefficient multiplying the square brackets expression in
eqs.(\ref{PAlfa}) and (\ref{PBeta}) is exactly $\frac{\pi}{8}
\frac{q}{r}k\eta_1$. Then, the final expressions for 
the $\alpha$ and $\beta$ coefficients are:
\bes
{\mid\alpha\mid}^2  
& = & \frac{\pi}{8}\;X\;
\left[ \left(1+\frac{{\left(\frac{q}{r}\right)}^2}{X^2}\right)
{{\mathcal{H}}_{\nu}}^{(1)}(X){{\mathcal{H}}_{\nu}}^{(2)}(X) \;+\;
\right. \nonumber \\
& + & {{\mathcal{H}}_{\nu-1}}^{(1)}(X){{\mathcal{H}}_{\nu-1}}^{(2)}(X) 
\nonumber \\
& - & \mbox{}\left. 2 \frac{\frac{q}{r}}{X}
{{\mathcal{H}}_{\nu}}^{(1)}(X){{\mathcal{H}}_{\nu-1}}^{(2)}(X)  
 - i \frac{4}{\pi } \frac{\frac{q}{r}}{X^2} + 
\frac{4}{\pi  X} \right] \label{PALFA}
\ees
\bes
{\mid\beta\mid}^2  
& = & \frac{\pi}{8} \; X \;
\left[ \left(1+\frac{{\left(\frac{q}{r}\right)}^2}{X^2}\right)
{{\mathcal{H}}_{\nu}}^{(1)}(X){{\mathcal{H}}_{\nu}}^{(2)}(X) \;+\;
\right. \nonumber \\
& + & {{\mathcal{H}}_{\nu-1}}^{(1)}(X){{\mathcal{H}}_{\nu-1}}^{(2)}(X) 
\nonumber \\
& - & \mbox{}\left. 2 \frac{\frac{q}{r}}{X}
{{\mathcal{H}}_{\nu}}^{(1)}(X){{\mathcal{H}}_{\nu-1}}^{(2)}(X)  
 - i \frac{4}{\pi } \frac{\frac{q}{r}}{X^2} - 
\frac{4}{\pi  X} \right] \label{PBETA}
\ees

\subsubsection{The Power Spectrum in the Full Dilaton Case}

\margen Using eq.(\ref{eq:pomega}) and (\ref{PBETA}), we have
for the power spectrum the next expression:
\bes \label{P:FD}
P(\omega) d\omega & = & \frac{\hbar}{8 \pi c^3}{\omega}^2  
\frac{{\left(\frac{q}{r}\right)}^2}{S} d\omega
\left[ \left(1+\frac{{\left(\omega \; S \right)}^2}
{{{\left(\frac{q}{r}\right)}^2}}\right)
{{\mathcal{H}}_{\nu}}^{(1)}(\omega S){{\mathcal{H}}_{\nu}}^{(2)}(\omega S) 
\;+\;
\right. \nonumber \\
& + & \frac{{\left(\omega \; S\right)}^2}
{{\left(\frac{q}{r}\right)}^2}
{{\mathcal{H}}_{\nu-1}}^{(1)}(\omega S){{\mathcal{H}}_{\nu-1}}^{(2)}(\omega S) 
\; + \; \nonumber \\
& - & \mbox{}\left. 2 \frac{\omega S}{\frac{q}{r}}
{{\mathcal{H}}_{\nu}}^{(1)}(\omega S){{\mathcal{H}}_{\nu-1}}^{(2)}(\omega S)  
 - i \frac{4}{\pi  \frac{q}{r}} - \frac{4}{\pi}\frac{\omega S}
{{\left(\frac{q}{r}\right)}^2} \right] 
\ees

The fraction of critical energy density takes the expression:
\bes \label{om:FD}
\Omega_{GW} & = & \frac{\hbar \; G}{3 {H_0}^2 c^5}\;
\frac{{\left(\frac{q}{r}\right)}^2}{S} \; \omega^3
\left[ \left(1+\frac{{\left(\omega \; S \right)}^2}
{{\left(\frac{q}{r}\right)}^2}\right)
{{\mathcal{H}}_{\nu}}^{(1)}(\omega S){{\mathcal{H}}_{\nu}}^{(2)}(\omega S) 
\;+\;
\right. \nonumber \\
& + & \frac{{\left(\omega \; S\right)}^2}
{{\left(\frac{q}{r}\right)}^2}
{{\mathcal{H}}_{\nu-1}}^{(1)}(\omega S){{\mathcal{H}}_{\nu-1}}^{(2)}(\omega S) 
\; + \;
\nonumber \\
& - & \mbox{}\left. 2 \frac{\omega S}{\frac{q}{r}}
{{\mathcal{H}}_{\nu}}^{(1)}(\omega S){{\mathcal{H}}_{\nu-1}}^{(2)}(\omega S)  
 - i \frac{4}{\pi  \frac{q}{r}} - \frac{4}{\pi}\frac{\omega S}
{{\left(\frac{q}{r}\right)}^2} \right] 
\ees

We compute for the three-dimensional case the values of the coefficients 
in front of the square brackets in eqs.(\ref{P:FD}) and (\ref{om:FD}).
These values differ from our previous computations on (\ref{cfp}) and 
(\ref{cfo}) 
in a factor ${\left(\frac{\frac{q}{r}}{\nu-\frac{1}{2}}
\right)}^2$. Thus, in the full dilaton case, we have:
\be \label{cfp2}
\frac{\hbar}{8 \pi c^3}\; \frac{{\left(\frac{q}{r}\right)}^2}{S} \: 
\sim \: 
4.008 \; 10^{-55} \; \frac{{\mathrm{erg}}\;{\mathrm{seg}}^3}{{\mathrm{cm}}^3}
\ee
\be \label{cfo2}
\frac{\hbar \; G}{3 {H_0}^2 c^5} \; \frac{{\left(\frac{q}{r}\right)}^2}
{S} \: \sim \:  6.220 \; 10^{-47} \; {\mathrm{seg}}^{-3}
\ee

In the Tables ({\ref{spectra}}) and ({{\ref{fracs}}}), 
we show the different
power spectra and fraction of critical energy density for
a single cosmological scale factor, but in the
three different roles for the dilaton.
As can be seen, there is 
a notorious difference between
the three different treatments of the dilaton role for a single 
cosmological scale 
factor description. First, the parameter $\nu$, characterizing the
inflationary expansion and the power spectrum, is different in each case.
The sign of this parameter is only ensured in the full dilaton treatment, 
by relating it with the form of the metric amplitude perturbation.
Second one, the coefficients in front of
the square brackets are different. Within the reduced amplitude
treatment (no dilaton and partial dilaton cases), these 
coefficients appear
related with the quantity $\left(\nu - \frac{1}{2}\right)$, 
meanwhile for the total metric amplitude $h_{\nu \mu}$ matching (full 
dilaton case), the 
coefficients have the expression $\left(\frac{q}{r}\right)$.

It must be observed that the No Dilaton Case is coherently 
described also by the expressions corresponding to the Full Dilaton Case.
In fact, in the first one, the next relation holds: 
\bef
\nu \; - \; \frac{1}{2} \; = \; \frac{q}{r}
\eef
(this relation is not satisfied by the partial dilaton role case, which 
must be described by the specific  expressions above indicated).

These observations enable us to work with general expressions for the
stochastic background of gravitational waves, given by the corresponding
ones of the full dilaton case eqs.(\ref{P:FD}) and (\ref{om:FD}).  The partial 
dilaton and no dilaton cases can be simply obtained from these
expressions by substituting the factors $\frac{q}{r}$ by 
$\left(\nu-\frac{1}{2}\right)$ with 
the proper value of parameter $\nu$ for each one.

\begin{table}[hp]
\centering 
\begin{tabular}{||c||c|c|c||l||}
\hline \hline
{\bf Case}& \multicolumn{2}{c|}{par. inflationary $\nu$} & {$(\nu-1)$}& 
{\it Power Spectrum} \\
$\empty$ & gral. & $d= 3$ & \empty & $P(\omega) d\omega$ \\ 
\hline \hline
{\bf No} & $\pm {\left(\frac{d-1}{2}q + \frac{1}{2}\right)}$ & $\frac{5}{6}$
 & $-\frac{1}{6}$ &
$ \frac{\hbar}{8 \pi c^3}{\omega}^2  
\frac{{\left(\nu-\frac{1}{2}\right)}^2}{S} d\omega$ \\
{\bf Dilaton} & \empty & \empty & \empty &
$ \left[ \left(1+\frac{(\omega S)^2}{{\left(\nu-\frac{1}{2}\right)}^2}\right)
{{\mathcal{H}}_{\nu}}^{(1)}(\omega S){{\mathcal{H}}_{\nu}}^{(2)}(\omega S) 
 \right. $ \\
\cline{1-4}
\cline{1-4}
{\bf Partial} & $\pm {\left(\frac{d+1}{2}q - \frac{1}{2}\right)}$ & 
$\frac{1}{6}$ & $-\frac{5}{6}$ &
$ \;+ \; \frac{(\omega S)^2}{{\left(\nu-\frac{1}{2}\right)}^2} \; 
{{\mathcal{H}}_{\nu-1}}^{(1)}(\omega S)
{{\mathcal{H}}_{\nu-1}}^{(2)}(\omega S) $ \\
{\bf Dilaton} & \empty & \empty & \empty &
$ \; - \;  \frac{2 \; \omega S}{\left(\nu-\frac{1}{2}\right)}
{{\mathcal{H}}_{\nu}}^{(1)}(\omega S){{\mathcal{H}}_{\nu-1}}^{(2)}(\omega S)$\\
\empty & \empty & \empty & \empty &
$ \; - \; \left. \frac{4\;\omega S}{\pi {\left(\nu-\frac{1}{2}\right)}^2}
\; - i \frac{4}{\pi \left(\nu-\frac{1}{2}\right)} \right]  $ \\
\hline \hline
{\bf Full} & ${\left(\frac{1}{2}\;-\; q \frac{d+1}{2}\right)}$ & 
$-\frac{1}{6}$ & $-\frac{7}{6}$ &
$\frac{\hbar}{8 \pi c^3}{\omega}^2  
\frac{{\left(\frac{q}{r}\right)}^2}{S} d\omega $ \\
{\bf Dilaton} & \empty & \empty & \empty &
$ \left[ \left(1+\frac{{\left(\omega \; S \right)}^2}
{{{\left(\frac{q}{r}\right)}^2}}\right)
{{\mathcal{H}}_{\nu}}^{(1)}(\omega S){{\mathcal{H}}_{\nu}}^{(2)}(\omega S) 
\right. $ \\
\empty & \empty & \empty & \empty &
$\; + \; \frac{{\left(\omega \; S\right)}^2}{{\left(\frac{q}{r}\right)}^2}
{{\mathcal{H}}_{\nu-1}}^{(1)}(\omega S){{\mathcal{H}}_{\nu-1}}^{(2)}(\omega S) 
$ \\
\empty & \empty & \empty & \empty &
$ \; - \;  2 \frac{\omega S}{\frac{q}{r}}
{{\mathcal{H}}_{\nu}}^{(1)}(\omega S){{\mathcal{H}}_{\nu-1}}^{(2)}(\omega S)
$ \\
\empty & \empty & \empty & \empty &
$ \; - \; \left. i \frac{4}{\pi  \frac{q}{r}} \; - \; 
\frac{4}{\pi}\frac{\omega S}
{{\left(\frac{q}{r}\right)}^2} \right]  $ \\
\hline \hline
\end{tabular}
\vspace*{7pt}
\caption{\label{spectra} The Dilaton Role in the Power Spectrum 
in String Cosmology}
\vspace{5pt}
{\parbox{135mm}{\footnotesize Comparative Table of the Gravitational Waves
Power Spectrum obtained in our String Cosmology Model. Differences have 
oughted to the weight of the dilaton in each case. The differences
between the three cases turn out in the value of the $\nu$
parameter and the coefficients in the power spectrum expresssions.}} 
\end{table}

\begin{table}[p]
\centering 
\begin{tabular}{||c||c|c|c||l||}
\hline \hline
{\bf Case}& \multicolumn{2}{c|}{par. inflationary $\nu$} & {$(\nu-1)$}& 
{\it Contribution to  Energy Density} \\
\empty & gral. & $d= 3$ & \empty & $\Omega_{GW}$ \\ 
\hline \hline
{\bf No} & $\pm {\left(\frac{d-1}{2}q + \frac{1}{2}\right)}$ & $\frac{5}{6}$
 & $-\frac{1}{6}$ &
$ \frac{\hbar \; G}{3 {H_0}^2 c^5}\;
\frac{{\left(\nu-\frac{1}{2}\right)}^2}{S} \; \omega^3$ \\
{\bf Dilaton} & \empty & \empty & \empty &
$ \left[ \left(1+\frac{(\omega S)^2}{{\left(\nu-\frac{1}{2}\right)}^2}\right)
{{\mathcal{H}}_{\nu}}^{(1)}(\omega S){{\mathcal{H}}_{\nu}}^{(2)}(\omega S) 
 \right. $ \\
\cline{1-4}
\cline{1-4}
{\bf Partial} & $\pm {\left(\frac{d+1}{2}q - \frac{1}{2}\right)}$ & 
$\frac{1}{6}$ & $-\frac{5}{6}$ &
$ \;+ \; \frac{(\omega S)^2}{{\left(\nu-\frac{1}{2}\right)}^2} \; 
{{\mathcal{H}}_{\nu-1}}^{(1)}(\omega S)
{{\mathcal{H}}_{\nu-1}}^{(2)}(\omega S) $ \\
{\bf Dilaton} & \empty & \empty & \empty &
$ \; - \;  \frac{2 \; \omega S}{\left(\nu-\frac{1}{2}\right)}
{{\mathcal{H}}_{\nu}}^{(1)}(\omega S){{\mathcal{H}}_{\nu-1}}^{(2)}(\omega S)
$ \\
\empty & \empty & \empty & \empty &
$ \; - \; \left. \frac{4\;\omega S}{\pi {\left(\nu-\frac{1}{2}\right)}^2}
\; - i \frac{4}{\pi \left(\nu-\frac{1}{2}\right)} \right]  $ \\
\hline \hline
{\bf Full} & ${\left(\frac{1}{2}\;-\; q \frac{d+1}{2}\right)}$ & 
$-\frac{1}{6}$ & $-\frac{7}{6}$ &
$ \frac{\hbar \; G}{3 {H_0}^2 c^5}\;
\frac{{\left(\frac{q}{r}\right)}^2}{ S} \; \omega^3 $ \\
{\bf Dilaton} & \empty & \empty & \empty &
$ \left[ \left(1+\frac{{\left(\omega \; S \right)}^2}
{{{\left(\frac{q}{r}\right)}^2}}\right)
{{\mathcal{H}}_{\nu}}^{(1)}(\omega S){{\mathcal{H}}_{\nu}}^{(2)}(\omega S) 
\right. $ \\
\empty & \empty & \empty & \empty &
$ \; + \; \frac{{\left(\omega \; S\right)}^2}{{\left(\frac{q}{r}\right)}^2}
{{\mathcal{H}}_{\nu-1}}^{(1)}(\omega S){{\mathcal{H}}_{\nu-1}}^{(2)}(\omega S) 
$ \\
\empty & \empty & \empty & \empty &
$ \; - \;  2 \frac{\omega S}{\frac{q}{r}}
{{\mathcal{H}}_{\nu}}^{(1)}(\omega S){{\mathcal{H}}_{\nu-1}}^{(2)}(\omega S)
$ \\
\empty & \empty & \empty & \empty &
$ \; - \; \left. i \frac{4}{\pi  \frac{q}{r}} \; - \; 
\frac{4}{\pi}\frac{\omega S}
{{\left(\frac{q}{r}\right)}^2} \right]  $ \\
\hline \hline
\end{tabular}
\vspace*{7pt}
\caption{\label{fracs} Contribution to Energy Density
with different Dilaton Role in String Cosmology}
\vspace{5pt}
{\parbox{135mm}{\footnotesize Comparative Table of the Fraction of the
Energy Density $\Omega_{GW}$ contained in Gravitational Wave Backgrounds 
produced in our
String Cosmology Model. Again, the differences  at the value of  $\nu$
parameter and have oughted to the weight of the 
dilaton role in each case. Signatures of this are the value of the $\nu$
parameter and the coefficients in the power spectrum expresssions.}} 
\end{table}

\begin{figure}[h]
\psfrag{0}{$1$}
\psfrag{3}{$10^{3}$}
\psfrag{4}{$10^{4}$}
\psfrag{6}{$10^{6}$}
\psfrag{5}{$10^{5}$}
\psfrag{7}{$10^{7}$}
\psfrag{8}{$10^{8}$}
\psfrag{9}{$10^{9}$}
\psfrag{12}{$10^{12}$}
\psfrag{j}[r]{$10^{-41}$}
\psfrag{i}[r]{$10^{-42}$}
\psfrag{P}{$P(\omega)$ $\left[\frac{{\mathrm{erg}}\;{\mathrm{seg}}}
{{\mathrm{cm}}^3}\right]$}
\psfrag{w}{$\omega\left[\frac{{\mathrm{rad}}}{{\mathrm{seg}}}\right]$}
\psfrag{no}{{\bf No Dilaton Case}}
\centering
\includegraphics[width=100mm]{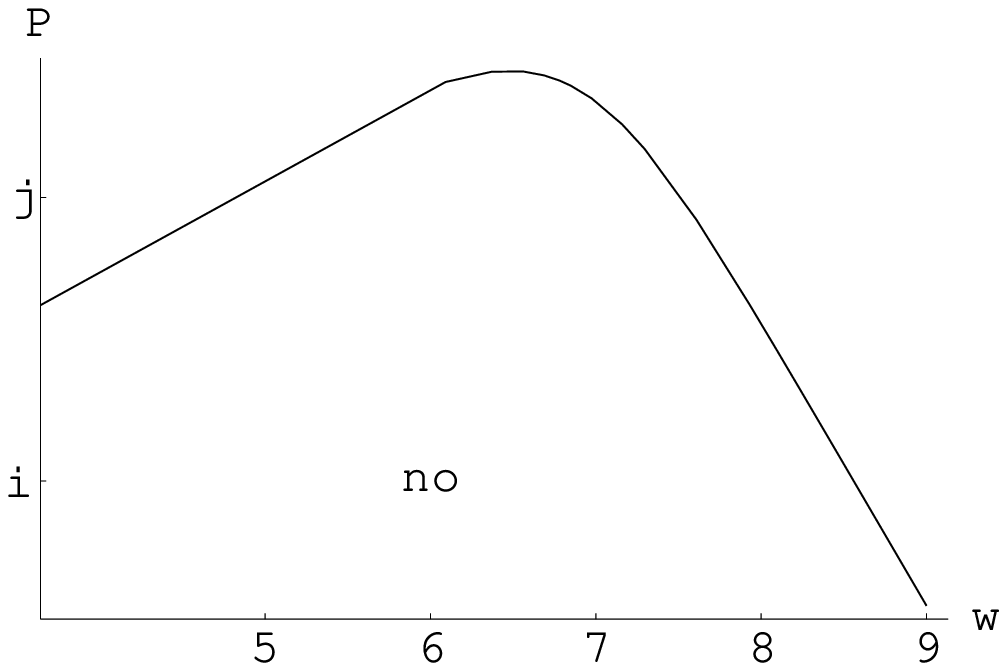}
\psfrag{i}[r]{$10^{-45}$}
\psfrag{j}[r]{$10^{-44}$}
\psfrag{k}[r]{$10^{-43}$}
\psfrag{l}[r]{$10^{-42}$}
\psfrag{si}{{\bf Partial Dilaton Case}}
\centering
\includegraphics[width=100mm]{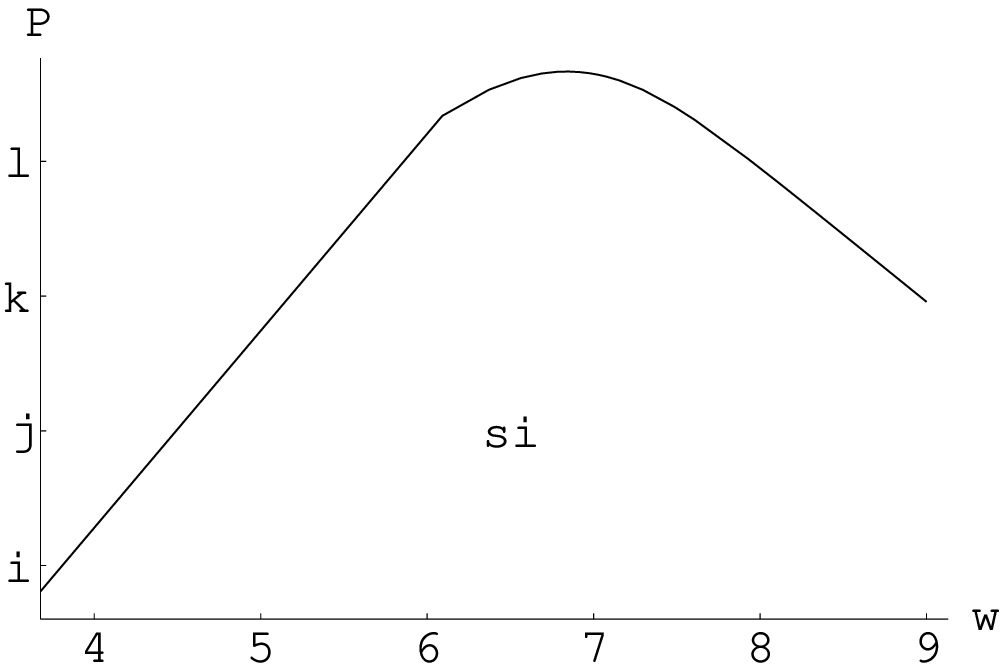}
\psfrag{j}[r]{$10^{-40}$}
\psfrag{k}[r]{$10^{-38}$}
\psfrag{l}[r]{$10^{-36}$}
\psfrag{m}[r]{$10^{-34}$}
\psfrag{si}{{\bf Full Dilaton Case}}
\centering
\includegraphics[width=100mm]{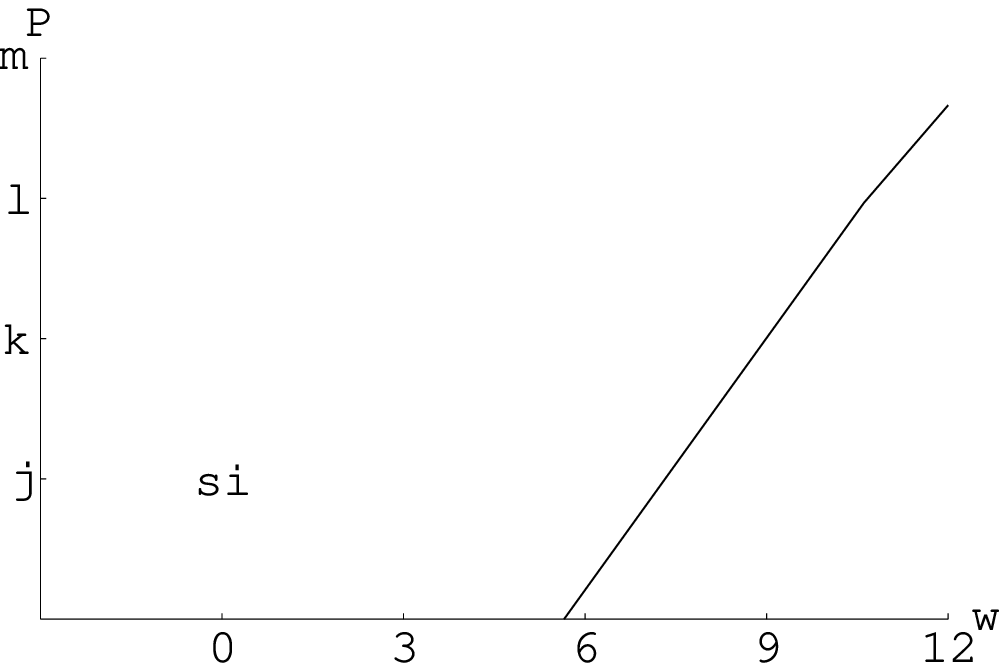}
\caption{\label{f.esp} Power spectrum for No Dilaton, partial Dilaton and 
Full Dilaton Case}
\end{figure}

\begin{figure}[h]
\psfrag{0}{$1$}
\psfrag{i}[r]{$10^{4}$}
\psfrag{3}{$10^{3}$}
\psfrag{6}{$10^{6}$}
\psfrag{7}{$10^{7}$}
\psfrag{8}{$10^{8}$}
\psfrag{9}{$10^{9}$}
\psfrag{12}{$10^{12}$}
\psfrag{j}[r]{$3\; 10^{-26}$}
\psfrag{l}[r]{$6\; 10^{-26}$}
\psfrag{m}{$\Omega_{GW}$}
\psfrag{w}{$\omega\left[\frac{{\mathrm{rad}}}{{\mathrm{seg}}}\right]$}
\psfrag{no}{{\bf No Dilaton Case}}
\centering
\includegraphics[width=100mm]{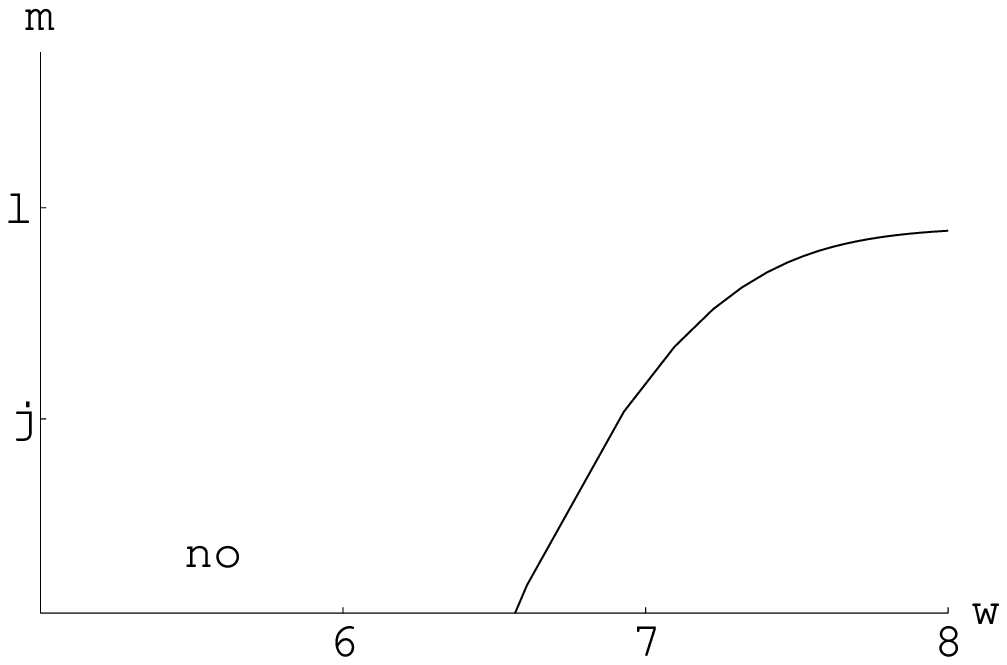}
\psfrag{i}[r]{$3 \;10^{-27}$}
\psfrag{j}[r]{$4 \; 10^{-27}$}
\psfrag{k}[r]{$6 \; 10^{-27}$}
\psfrag{n}[r]{$1 \; 10^{-26}$}
\psfrag{o}[r]{$1.4 \; 10^{-26}$}
\psfrag{p}[r]{$2 \; 10^{-26}$}
\psfrag{si}{{\bf Partial Dilaton Case}}
\centering
\includegraphics[width=100mm]{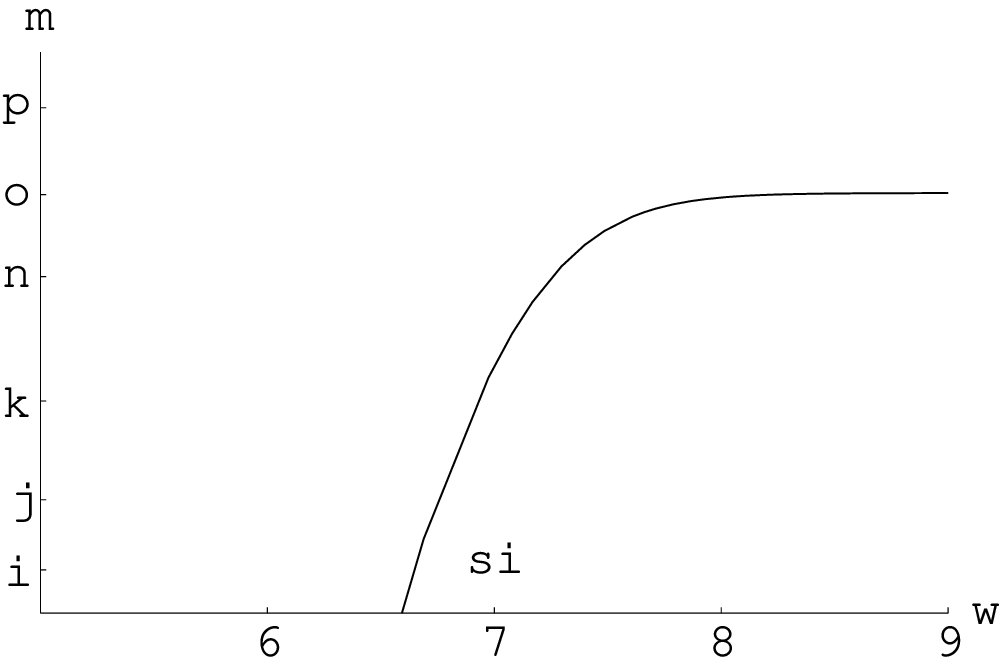}
\psfrag{i}[r]{$10^{-25}$}
\psfrag{j}[r]{$10^{-21}$}
\psfrag{k}[r]{$10^{-17}$}
\psfrag{l}[r]{$10^{-13}$}
\psfrag{si}{{\bf Full Dilaton Case}}
\centering
\includegraphics[width=100mm]{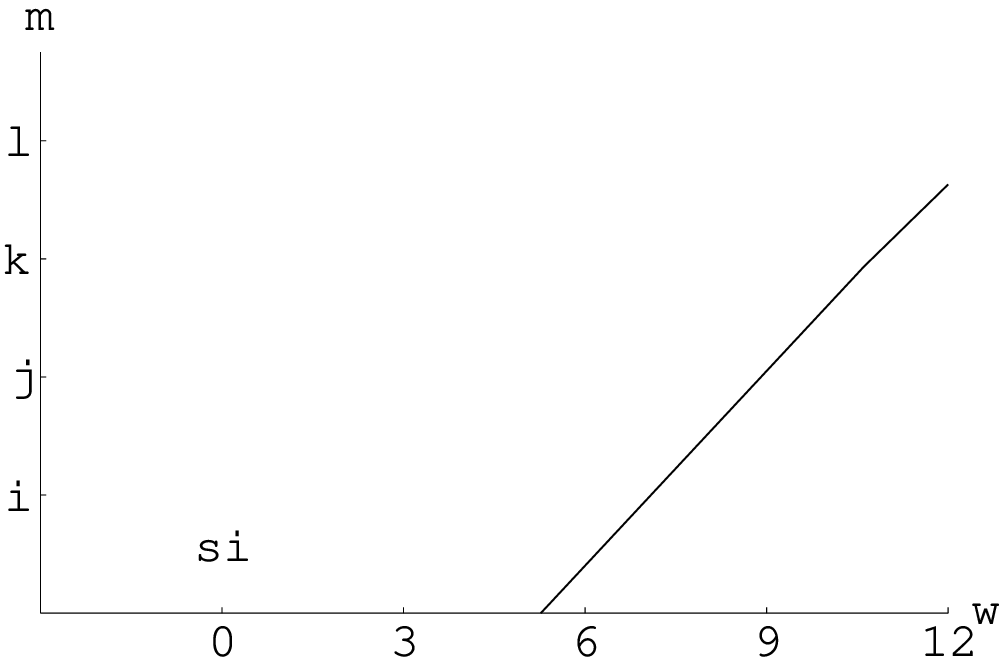}
\caption{\label{f.ome} Contribution to Energy Density for the No Dilaton, 
partial Dilaton and Full Dilaton Case}
\end{figure}
\clearpage

\section{Asymptotic Behaviours of the Gravitational Wave Background}
\setcounter{equation}0

\margen We have computed the analytical expressions for 
$P(\omega)d\omega$ and $\Omega_{GW}$ at
high and low frequencies. Both
regimes correspond to very large ($X>>1$) and very small ($X<<1$)
arguments of Hankel functions, respectively.
We consider the three regimes:
low frequencies for $\omega << \omega_x $ ($X<<1$), 
medium frequencies for $\omega \sim 1$
(exact analytical expressions 
(\ref{P:All}),(\ref{om:All}),(\ref{P:FD}),(\ref{om:FD})) and
high frequencies for $\omega >> \omega_x$ ($X>>1$)
where $\omega_x$ is the characteristic frequency (eq.(\ref{omx}))
with order of Mhz for the scale factor evolution considered.

The exact analytical form of the power spectrum, eqs.(\ref{P:All}) and
(\ref{P:FD}) requires a careful algebraic treatment in order to compute
their asymptotic behaviours.
Attention must be put both in the 
range of validity as on the order of expansions used.
Computation is quite long, thus we omit details here and give only
the leader orders. 

\subsection{High frequencies}
\enlargethispage{8mm}

\margen The High Frequency Asymptotic expressions for the No Dilaton $(ND)$ and
Partial Dilaton $(PD)$ Cases are formally equivalent. The leader orders for
the power spectrum and contribution to energy density can be written 
respectively as:
\bes \label{as.hnd}
\left. P(\omega) d\omega \right|_{ND,PD} & \sim & \frac{\hbar}{16 \pi^2 c^3}
\frac{{\left(\nu-\frac{1}{2}\right)}^2{\left(\nu+\frac{1}{2}\right)}^2}{S^4}
\omega^{-1} + O({\omega^{-3}}) \\
\left. \Omega_{GW} \right|_{ND,PD} & \sim & \frac{\hbar \; G}
{6 {H_0}^2 \pi c^5}
\frac{{\left(\nu-\frac{1}{2}\right)}^2{\left(\nu+\frac{1}{2}\right)}^2}
{S^4} + O({\omega^{-2}}) \label{aom.hnd}
\ees
\margen The Full Dilaton $(FD)$ Case presents different
expressions for the High Asymptotic Behaviour:
\bes \label{as.hfd}
\left. P(\omega) d\omega \right|_{FD} & \sim & \frac{\hbar}{4 \pi^2 c^3}
\frac{{\left(\frac{q}{r}\right)}^2}{S^2}
\omega \; {\left(\frac{\nu-\frac{1}{2}}{(\frac{q}{r})}
\;-\;1\right)}^2 + O({\omega^{-1}}) \\
\left. \Omega_{GW} \right|_{FD}& \sim & \frac{2 \hbar \; G}{3 {H_0}^2\pi c^5}
\frac{{\left(\frac{q}{r}\right)}^2}{S^2}
\omega^2 \;{\left(\frac{\nu-\frac{1}{2}}{(\frac{q}{r})}
\;-\;1\right)}^2 + O({\omega^{0}}) \label{aom.hfd}
\ees
\margen Notice that leader order terms have fixed dependences, independent of
number of spatial dimensions or parameters of the solutions. In fact, 
$P(\omega)d\omega$ in both No Dilaton and Partial Dilaton cases present 
a convergent behaviour at high frequencies as $\sim {\omega}^{-1}$.
In contrast, the Full Dilaton case is divergent in this regime
as $\sim {\omega}^{1}$. $\Omega_{GW}$ in the No Dilaton
and Partial Dilaton cases reachs asymptotically a constant value.
In the Full Dilaton case $\Omega_{Gw}$ diverges as $\omega^{2}$.
 
\subsection{Low frequencies}
\enlargethispage{9mm}

{\margen} Opposite to the high frequencies case,
leader orders for low frequencies depend explicitly from the parameter $\nu$.
As a consequence, each case presents a particular expression for
their asymptotic behaviours.
To leader order uniquely, the No Dilaton case and the Partial Dilaton
case have the same formal expressions:
\bes \label{Pl:ND}
\left. P(\omega) d\omega \right|_{ND, PD}& \sim & \frac{\hbar}{8 \pi c^3}
\frac{{\left(\nu-\frac{1}{2}\right)}^2}{S^3} d\omega
\frac{2^{2 \nu}}{\pi^2}{\Gamma(\nu)}^2{(\omega S)}^{2-2\nu}
\;+\;O((\omega S)^{2}) \\
\left. \Omega_{GW}\right|_{ND, PD} & \sim & \frac{\hbar \; G}{3 {H_0}^2 c^5}\;
\frac{{\left(\nu-\frac{1}{2}\right)}^2}{ S^4} \; 
\frac{2^{2 \nu}}{\pi^2}{\Gamma(\nu)}^2{(\omega S)}^{3-2\nu}
\;+\;O((\omega S)^{3}) \label{Ol:ND}
\ees

In the three-dimensional case, $\nu=\frac{5}{6}$ for the $(ND)$ case
and $\nu=\frac{1}{6}$ for the $(PD)$ case. We have 
at very low frequencies ${\left. P(\omega) d\omega\right|}_{ND} 
\sim \omega^{\frac{1}{3}}$
and ${\left. \Omega_{GW}\right|}_{ND} \sim \omega^{\frac{4}{3}}$,
${\left. P(\omega) d\omega \right|}_{PD} \sim 
\omega^{\frac{5}{3}}$ and ${\left. \Omega_{GW}\right|}_{PD} \sim 
\omega^{\frac{8}{3}}$.

In the Full Dilaton Case the low frequencies behaviour is:
\bes \label{Pl:FD}
\left. P(\omega) d\omega \right|_{FD}  \sim  \frac{\hbar}{8 \pi c^3}
\frac{{\left(\frac{q}{r}\right)}^2}{S^3} d\omega
\frac{{\Gamma(1-\nu)}^2}{2^{2 \nu}{\pi}^2}{\left(\frac{2}{\frac{q}{r}}
-\frac{1}{\nu}\right)}^2 {(\omega S)}^{2 + 2 \nu}+  
O((\omega S)^{2}) & &\\
\left. \Omega_{GW} \right|_{FD} \sim \frac{\hbar \; G}{3 {H_0}^2 c^5}\;
\frac{{\left(\frac{q}{r}\right)}^2}{S^4} \; 
\frac{{\Gamma(1-\nu)}^2}{2^{2 \nu}{\pi}^2}{\left(\frac{2}{\frac{q}{r}}
-\frac{1}{\nu}\right)}^2 {(\omega S)}^{3 + 2 \nu} +
O((\omega S)^{3}) & & \label{Ol:FD}
\ees

The leader order has the same behaviour $P(\omega) d\omega \sim 
\omega^{\frac{5}{3}}$ and $\Omega_{GW} \sim \omega^{\frac{8}{3}}$ as
in the Partial Dilaton Case, but the coefficient
is different.

The leader order at very asymptotic behaviours for all cases is
summarized in table (\ref{limi}). 
 
\begin{table}[hb]
\centering 
\begin{tabular}{||c|c||c|c||c|c|| }
\hline \hline
\multicolumn{2}{||c||}{\bf  Case} & \multicolumn{2}{c||}{$P(\omega)d\omega$} &
\multicolumn{2}{c||}{$\Omega_{GW}$} \\
\cline{3-6} \cline{3-6}
\multicolumn{2}{||c||}\empty  & 
{\bf Low Frq.} & {\bf High Frq.} &
{\bf Low Frq.} & {\bf High Frq.} \\
\hline \hline
{\it String}  & No Dilaton & $\sim \omega^{\frac{1}{3}}$ &
$\sim \omega^{-1}$ & $\sim \omega^{\frac{4}{3}}$ & cte. \\
\cline{2-6}
{\it Driven} & partial Dil. & $\sim \omega^{\frac{5}{3}}$ &
$\sim \omega^{-1}$ & $\sim \omega^{\frac{8}{3}}$ & cte. \\
\cline{2-6}
\mbox{\empty}  & Full Dil. & $\sim \omega^{\frac{5}{3}}$ &
$\sim \omega $ & $ \sim \omega^{\frac{8}{3}}$ & $\sim \omega^2 $   \\
\hline \hline
\end{tabular}
\vspace*{5pt}
\caption{\label{limi} Asymptotic Behaviours of  Gravitational Wave Power
Spectrum and Contribution to Energy Density.}
\vspace{5pt}
{\parbox{135mm}{\footnotesize The dominant dependence for very low
and very high frequences are summarized. The values are refered to the
three spatial dimensional case.}} 
\end{table}

\section{Discussion and Conclusions}
\setcounter{equation}0

\margen We have studied the production of a primordial stochastic gravitational
wave background in a cosmological model fully extracted in the context
of selfconsistent string cosmology. As seen in Section II,
suitable descriptions for inflationary, radiation
dominated and matter dominated stages can be extracted from effective 
string theory
in which strings propagating in the curved backgrounds are the
classical sources of matter and drive selfconsistently the cosmological
background.
The equation of state provided by the dynamics of string matter 
naturally generates in its evolution the
background in the three stages.

No exact description of the transitions among stages can be extracted
in the framework of this effective treatment. Higher order
corrections in low energy effective equations or more detailed
study on the evolution of the gas of strings would be 
neccessaries in order to handle such subject. 

This lack of full understanding of the transition dynamics is an
usual limitation when primordial metric perturbations are
computed.  The main consequence is the loss of predictibility.
Usual gravitational wave results depend
on parameters of the cosmological model which are not always well
known (\cite{gv93},\cite{buon}), or whose physical meaning must
be recovered  after computation (\cite{bggv}), or where the not
considered transition dynamics must be recalled in order to
overcome divergences (\cite{A}, \cite{bgv}). Usual arguments
dealing with horizon exit and reentry (\cite{gks},\cite{gkprd},\cite{Sah})
require a suitable cosmological description in order to be
translated univocally into current measurement abilities.

We faced this problem before to compute the metric
perturbations.  We elaborate a cosmological
minimal model of evolution for the scale factor with satisfactory 
sudden and continuous transitions, provided it makes use of particular
descriptive cosmic time variables at each stage (see eqs.(\ref{Descr})).
In this way, we merge our lack of knowledge about real 
transitions by modelizing them in a descriptive scale factor expressed
in suitable cosmic time variables. This description
is linked with the minimal observational Universe information
(standard values for transition times and scale factor ratii
reached in each stage). It provides an equivalent evolution for
the scale factor, but evades the undetermined region of transitions
without introduction of free parameters or loss of predictibility.
Unless the details imprinted  by dynamics transitions, in our case
modelized as sudden and continuous, our computation is expected
equivalent to one made on a full physical model running on cosmic time. 

We have also imposed continuity and smothness on the conformal time 
description, by constructing this on the descriptive variables
(see eqs.(\ref{EDescr})).
Continuity of scale factor both in cosmic time and conformal
time description is a feature not always present in literature (see \cite{p1}).
In our case, the variables constructed enable us to maintain
the linking with the observational Universe information. This
model seems us more appropiated in order to perform primordial
metric perturbations computations, since no effects oughted
to discontinuity in the scale factor of the metric are introduced.
Not so perfect is the treatment of the dilaton field, whose
evolution can be treated solely continuous but not smooth.
Further study would be needed in order to valutate this effect.

The only gravitational waves we have computed are those produced
as amplification by the evolution of background of tensorial metric 
perturbations. We do not account here another possible mechanisms of generation
neither amplification of another metric perturbations.

The variable $X$ in the power spectrum and the proper frequency
$\omega$ are related in a way totally determined by the description of 
the cosmological scale factor evolution, see eq.(\ref{ese}). 
The factor relating them depends 
of the expansion ratii, the exit time of inflationary 
epoch and the coefficients of expansion inflationary and current epochs.
Being all them fixed in our cosmological background linked to the
observational Universe, no free parameters are
introduced at this level. 
No one of remaining unknown parameters, like
global scale factor ${\bar{A_{II}}}$, appears on the results of
our computation.
Differently from almost all string cosmology 
computations in literature, firm predictions on precise frequencies ranges
can be extracted in our case. 

In this way, we have computed exact, fully predictive and free-parameter 
expressions for the power spectrum 
$P(\omega) d\omega$ and contribution to energy density $\Omega_{GW}$ 
of the primordial gravitational waves background
generated in the transition among the inflationary and radiation dominated
stages. We have not considered the contribution given by the radiation
dominated-matter dominated transition in this study. The gravitational wave
contribution due to this transition is expected to be neglectelly small,
as compared to the first transition. It would be expected the second
transition having a role only on the low 
frequencies regime \cite{A}, not so important in anycase for
our results.

\subsection{The Dilaton Role}
\enlargethispage{5mm}

\margen We have obtained drastic differences in the stochastic background of
gravitational waves produced in the same scale factor evolution by considering 
differents degrees of the role played by the dilaton. The simplest case, 
without account of the effect
of the dilaton neither on the perturbation equation nor on the amplitude 
perturbation.
The second case, a partial account, with the proper perturbation equation but
still matching the reduced amplitude perturbation. The lastest case, 
a full account 
by working with the total tensorial amplitude perturbation and 
perturbation equation.

The background of gravitational waves comes characterized in their shape 
by the parameter $\nu$, which depends of the inflationary description, 
the inflation-radiation dominated transition (the only on which we are
computing gravitational
wave production) and the role played by the dilaton. 
The expressions for $\nu$ have been found in the three cases
(eqs.(\ref{nu:ND}),(\ref{nu:SD}) and (\ref{nu:FD2})). 
We obtain an exact expression for the
power spectrum and energy density contribution (eqs.(\ref{P:All})
and (\ref{om:All})) in terms of Hankel functions of order $\nu$, formally 
equal in the No Dilaton and partial Dilaton cases. Differences among
them are oughted to $\nu$ differences. 
The formal expressions in the full dilaton case are 
different (eqs.(\ref{P:FD}),(\ref{om:FD})) both in  parameter $\nu$
as in coefficients involved.

The low frequency
and high frequency asymptotic regimes have been discussed.
In the No Dilaton case, asymptotic behaviours for power spectrum are
both vanishing at low and high frequencies as
$\omega^{\frac{1}{3}}$ and ${\omega}^{-1}$ respectively. 
This gives a gravitational wave contribution to the energy density 
asymptotically constant at high
frequencies of magnitude $\Omega_{GW} \sim 10^{-26}$.
There is a slope change that 
produces a peak in the power spectra around a characteristic frequency totally
determined by the model of 
$\omega_x \; \sim  \: 1.48 \; {\mathrm{Mhz}}$.

The Partial Dilaton case introduces the effect of the dilaton exclusively
in the tensorial perturbation equation (which is not longer
equivalent to the massless real scalar field propagation equation
\cite{fpa},\cite{wh}), but not on the perturbation itself. The general
characteristic are very similar to the previous case.
Both asymptotic regimes for $P(\omega)d\omega$ vanish again, but with 
dependences ${\omega}^{\frac{5}{3}}$ and ${\omega}^{-1}$.
The peak appears around the same characteristic frequency,
with value one order of magnitude lower than in the
No Dilaton case, as well as the asymptotic constant contribution
to energy density.

In contrast, when the full
dilaton role is accounted, general characteristics as well as orders
of magnitude of the spectrum are drastically modified.
It has similar values for the frequencies
below the Mhz, with power spectrum vanishing again as
$\omega^{\frac{5}{3}}$. For high frequencies, in opposition to the 
former cases, both $P(\omega) d\omega$ and $\Omega_{GW}$ are
increasing at high frequencies.
For $P(\omega)d\omega$, 
an asymptotic divergent behaviour proportional 
to $\omega$ is found. It gives values
much higher than the no dilaton and partial dilaton cases. 
The contribution to
$\Omega_{GW}$ is equally
divergent at high frequencies as $\omega^{2}$. The change of slope is
less visible and no clear peaks are found. The transition
from the low frequency to the high frequency regime is 
slower than in the previous case and the full analytical
expressions are needed on a wider range 
$10^6 \sim 10^9 Hz$. 
Comparative tables of these summarized 
results can be
found in tables (\ref{spectra}), (\ref{fracs}) and (\ref{limi}).
See also figs.(\ref{f.esp}) and (\ref{f.ome}).

Existence of an upper cutoff must be studied, whereas the Full
Dilaton Case studied could be in contradiction with 
observational bounds as the current value of total
energy density in critical units $\Omega \sim 1$.
In that case, an end-point not predicted by the current
minimal model considerations
could be introduced in the spectrum as made in the literature \cite{bgv}.
Divergent high frequencies behaviour and introduction of an upper
cutoff is an usual feature in the string cosmology contexts.

\subsection{String and No-String  Cosmologies}

\margen Among the spectra computed in string cosmology contexts,
it must be distinguished between those computed in  
Brans-Dicke frames (that we compare with our Full Dilaton
Case) and those computed following an usual quantum field theory way,
that is, as our No Dilaton Case. 

The shape of the spectra computed in 
string cosmology contexts are very 
similar. The principal features, as slope changings, are signal
of the number of stages or transitions considered in the scale
factor evolution. All the known cases, coherently treated in
Brans-Dicke frames, presents an increasing dependence at
high frequencies. 

Notice that we are not handling with a ``Pre-Big Bang'' scenario \cite{gv93}. 
Our String Driven Cosmological Background
runs on positive values for the proper cosmic time $t$.
It have been said \cite{bgv} that Pre Big-Bang scenario predicts 
a power spectrum with a peak and an end-point. Our analysis
enable us to identify these features with the Brans-Dicke
framework where the low effective energy treatment is
made, without necessity of a ``Pre Big Bang'' phase. 
In fact we have seen as the same Universe evolution
and transitions have given different power spectra. In
the full dilaton treatment, the transition among the inflationary
inverse power stage and the radiation dominated stage gives
an always increasing function for the number of particles
${\mid \beta \mid}^2$. It causes the high frequency range 
to be asymptotically divergent, feature that disappears completely
in the No Dilaton or in the Partial dilaton treatments. 

Similarly, a cosmological model
involving successively a named Dilaton Driven stage plus an ``String Phase'' at
nearly constant curvature, and a radiation dominated stage produces 
also a spectrum 
increasing with frequency (\cite{bggv}). 
As made in the literature (\cite{bgv}), an upper
cutoff is placed on this increasing spectra by considering the frequency 
making ${\mid \beta(\omega) \mid}^2 = 1$ as the maximum one
produced.  This frequency is the one where 
one graviton per space phase unit volume and polarization is produced,
and after which exponential suppression is supposed. These arguments
leds to identification of the so called ``end-point'' that acts as ``peak'' of
the spectrum oughted
to the increasing power with frequency. The exact
values at current time are functions of the undetermined parameters in
the cosmological model there treated \cite{bggv}.
\pagebreak[4]

We conclude that whatever gravitational wave computation on inflationary stages
of the type extracted in string cosmology, coherently made in the
Brans-Dicke frame, must give an increasing spectrum.
The peaks are produced by slope change and  they are signal of the 
transitions in
the dynamics of background evolution. 
We consider the Pre-Big Bang scenario do not predicts
a peak, but it is supposed by defining a $\omega_1$ such that
$|\beta(\omega_1)|^2=1$. This proper frequency 
is computed at the beginning of radiation dominated stage, when
the wave reenters the horizon and it must suffer a
redshift at current time, expressable as function of unknown parameters
of a ``string phase''. It acts as 
end-point because waves with $\omega>\omega_1$
are supposed exponentially suppressed. Since the spectrum was increasing
with frequency, the same frequency $\omega_1$ constitutes 
a maximun (peak).

If we use the same argument in order to fix an upper limit,
our spectrum must be cutted at frequency 
$\omega_{max} \sim 3.85$ MHz 
where the power spectrum will have a value around 
$ P(\omega) d\omega \sim 5.68 \; 10^{-41} \frac{\mathrm{erg.s}}
{{\mathrm{cm}}^3}$
and $\Omega_{GW} \sim 3.40 \; 10^{-26} \rho_{c}$.
These are the same order of magnitudes of the peaks atteinted on the
No Dilaton and partial Dilaton Cases. No conflict with observational
constraints look possible for such weak signals predicted. But this
argument could be too naive, since in the practical way is equivalent
to cancelate from the spectra the features introduced by the full
dilaton role. 

For instance, 
if we want to relaxe this condition and we ask us what would
be the gravitational wave with maximum possible frequency,
we can give at least a coherency condition.  Now, let us ask if it
is possible to produce a wave with period lower than the
Planck time, that means with wavelenght lower than the Planck
lenght scale. If that feature is possible, it is clear that our
simplified quantum theory formalism would be unappropiated
in order to compute such produced particles. We can fix the
upper cutoff in the corresponding $\omega_{P} \sim 10^{44} Hz$ 
by considering  that beyond it the power spectrum of gravitational waves must
be, at least, different from elsewhere hitohere computed. At those
values, both the No Dilaton and partial Dilaton case predicts
completely negligible production, but not is the case for Full
Dilaton Case. Its asymptotic divergent behaviour predicts
enormous values for 
$ P(\omega) d\omega$ and $\Omega_{GW}$. 

Obviously, we do not mean that these are the right predictions of
string cosmology. Without doubt, the model here treated is 
excessively simple in order to handle properly the end-point
of spectrum. But usual assumptions on upper exponential
suppression could be more careful revisited, since a too naive
treatment in this point can mask important features of predictions.

In relation with  no-string inflationary cosmologies, the so called 
standard inflation is usually intended as a
De Sitter stage. Notice that string driven inflationary stage describes an 
evolution with inverse power dependence. It must not be confused
with usual power law, although our model can be said superinflationary
too.

There is a radical difference among the string cosmology spectra
and those obtained with an exponential inflationary expansion
\cite{A}. The divergence
at low frequencies that it supposes is not found in
our String Driven Cosmological Background 
The high frequencies behaviour are compatible with the No Dilaton
Case (\ref{P:All}), computed in a totally equivalent way. 

If comparaison is made with the Full Dilaton Case, a totally different
behaviour is found with respect to the obtained in De Sitter case.
Notice that comparaison among  string comology 
inflationary models and standard inflation means confrontate
power and inverse power type laws with De Sitter exponential
inflation, since until the moment no De Sitter type expansion have
been coherently obtained in String Cosmology. This 
difference in the inflationary scale factor dynamics, together with 
the appropiated teatment of metric perturbations in each framework
are the main causes of differences
among the gravitational wave power spectra in both cases.

No explicit dependence on the beginning of the inflationary stage $t_i$
has been found on the gravitational wave computation.
In anycase, further
study must be done in order to determine the influence on the
power spectra if earliers inflation stages are considered.
Gravitational wave production in frameworks with multistages
inflation can be found  in literature, both in the most classical
cases (\cite{gksqz},\cite{Sah},\cite{wh}) 
meanwhile in the string cosmology contexts there are various
multistage models (\cite{gv93},\cite{bggv},\cite{buon}).

In summary, there are a variety of effects affecting the gravitational wave 
power spectra and justifiying the differences among our results
and another string and no string cosmologies.
The most notorious difference is 
oughted to the existence of a dilaton field and a metric evolving in a 
Brans-Dicke frame. Differences are found on the equation for the tensorial 
perturbation, and the amplitude of perturbation depends itself on the
dilaton field.
Our spectra here computed have showed clearly the successive effect obtained 
on the gravitational wave production by introducing the dilaton effect on the
perturbation equation (partial Dilaton case) and on the perturbation itself 
(Full Dilaton case). From an unique evolution of the scale factor, 
we have seen 
as the dilaton role on the spectra consists in increasing the high
frequencies regime, leading to a divergent behaviour.
This is in agreement with almost string cosmology,
Brans-Dicke computations.
In our cases without full account of the dilaton role, we have spectra
decreasing at higher frequencies, as such obtained in the
literature \cite{A} in the more standard frameworks.
\enlargethispage{5mm}

There are many points that desserve further study in order to
be clarified, either from the point of view of  string cosmological
backgrounds and transitions dynamics, either from those of
gravitational wave
computations in the string appropiated frameworks. 
Better treatments than here applied could take place
in every phase of the problem, advising us against consider
this procedure or its results as definitives. 
In anycase,
we have proved here some aspects of the way predictions on
observational consequences can be extracted from String Cosmology. 

{\bf ACKNOWLEDGMENTS}

M.P.I. thanks Fundaci\'on Empresa Universidad de Zaragoza 
and Chambre \linebreak[4] R\'egionale de Commerce et d'Industrie d'\^Ile-de-France 
for support under C.E.E.
Leonardo Programme during this work.


\end{document}